\documentclass[12pt]{article}
\pdfoutput=1
\usepackage{amsmath,amssymb,amsfonts,amsthm,bm,bbm,cancel,wasysym}
\usepackage{epsfig,graphics,graphicx,epstopdf}
\usepackage{array,booktabs,colortbl,colordvi,multirow}
\usepackage{colordvi,color,xcolor}
\usepackage{hyperref}
\usepackage{rotating}

\usepackage{verbatim}
\usepackage{cite}
\usepackage{subfig}
\usepackage{setspace}
\usepackage{url}
\usepackage[percent]{overpic}
\usepackage{slashed}
\usepackage{authblk}
\usepackage{xspace}
\usepackage{fullpage}
\usepackage{hyperref}

\def\ie{{\it i.e.}}
\def\eg{{\it e.g.}}
\def\etc{{\it etc}}

\def\to{\rightarrow}

\newskip\zatskip \zatskip=0pt plus0pt minus0pt
\def\matth{\mathsurround=0pt}
\def\lsim{\mathrel{\mathpalette\atversim<}}
\def\gsim{\mathrel{\mathpalette\atversim>}}
\def\atversim#1#2{\lower0.7ex\vbox{\baselineskip\zatskip\lineskip\zatskip
  \lineskiplimit 0pt\ialign{$\matth#1\hfil##\hfil$\crcr#2\crcr\sim\crcr}}}

\begin{document}


\begin{flushright}
SLAC-PUB-17178\\
\today
\end{flushright}
\vspace*{5mm}

\renewcommand{\thefootnote}{\fnsymbol{footnote}}
\setcounter{footnote}{1}

\begin{center}

{\Large {\bf Kinetic Mixing, Dark Photons and an Extra Dimension: I}}\\

\vspace*{0.75cm}

{\bf Thomas G. Rizzo}~\footnote{rizzo@slac.stanford.edu}

\vspace{0.5cm}

{SLAC National Accelerator Laboratory}\ 
{2575 Sand Hill Rd., Menlo Park, CA, 94025 USA}

\end{center}
\vspace{.5cm}

\begin{abstract}
 
\noindent

Extra dimensions (ED) can provide a useful tool for model-building. In this paper we introduce a single, flat ED extension of the kinetic-mixing/dark photon (DP) portal  for dark matter (DM) 
interactions with the Standard Model (SM) assuming a compactification `radius' of order $R^{-1}\sim10-1000$ MeV and examine the resulting modifications to and augmentation of the 
usual DP phenomenology. In the present scenario, both the DP and DM experience the full 5-D while the SM fields are constrained to lie on a 4-D brane at the boundary of the ED. Such a 
setup can naturally yield the observed value of the DM relic density and explain the required rough degeneracy of the DM and DP masses needed to obtain it. Gauge symmetry breaking can 
occur via boundary conditions without the introduction of an additional singlet Higgs scalar thus avoiding all constraints associated with the coupling of such a field to the usual SM Higgs 
field in 5-D. The self-consistency in the removal of the kinetic mixing terms is found to lead to a brane localized kinetic term for the 5-D gauge field on the SM brane.  Multiple variations of 
this scenario are found to be possible which are consistent with current experimental constraints but which predict very different phenomenologies. In this paper,  we discuss the case of 
a complex scalar 5-D DM field, consistent with constraints arising from the CMB, which may or may not obtain a vacuum expectation value (vev). This approach can lead to interesting and 
distinctive signatures while being constrained by a wide array of existing measurements but with the details being dependent upon the model specifics.
\end{abstract}

\renewcommand{\thefootnote}{\arabic{footnote}}
\setcounter{footnote}{0}
\thispagestyle{empty}
\vfill
\newpage
\setcounter{page}{1}



\section{Introduction}

The nature of Dark Matter (DM) is one of the greatest mysteries in particle physics: the Standard Model (SM) provides us with no candidates for such particles and 
forces us to entertain new physics scenarios for a possible explanation.  Weakly Interacting Massive Particles (WIMPS)\cite{Arcadi:2017kky},  particularly in the 
form of the Higgsino/gaugino Supersymmetric partners (as well as axions\cite{Kawasaki:2013ae,Graham:2015ouw}) have generally been the most popular of the DM 
candidates. One reason for this is that the UV complete scenarios wherein such particles might arise were motivated to solve other problems and the existence of a potential DM 
candidate within them was a welcome bonus. However, 
as the predicted WIMP signatures have failed to show themselves at the LHC or in either direct (DD) or indirect detection (ID) experiments\cite {susy17}, it 
behooves us to widen our theoretical viewpoint as well as our experimental  search windows. This is particularly true for the case of lighter DM masses, below that of the 
traditional WIMP mass scale, where many of the conventional searches clearly falter. This is the strong message contained in the white papers from both the Dark 
Sectors Workshop\cite{Alexander:2016aln} and the U.S. Cosmic Visions Workshop\cite{Battaglieri:2017aum}. Of course, without the guidance of more UV complete 
theories, such as SUSY, it is difficult to focus, {\it a priori}, on any  particular mass range without some input from experiment as to where we can or should look given our existing 
and potential future capabilities. One possibility is, roughly,  the $\sim$10 MeV to 1 GeV mass range where the standard WIMP-targeted DD experiments involving 
nuclear targets are found to suffer due to detector energy thresholds as well as from potentially serious neutrino backgrounds but which can be accessed 
by other means\cite{Battaglieri:2017aum}. In this mass range the DM can still be in thermal contact with the SM in the early universe and a modified version of the WIMP paradigm 
associated with thermal relic freeze-out can still go through, albeit with new, non-SM interactions responsible for achieving the observed relic density.
 
Theories employing extra spatial dimensions (ED) at the $\sim$ TeV scale\cite{ED} have provided us with very useful tools for both model building and as means  to 
attempt addressing the outstanding issues within the SM, in particular, the gauge hierarchy\cite {ED} and flavor problems\cite {flavor}. Such theories can also provide 
a potential origin for a TeV scale, non-SUSY version of WIMP DM\cite {ued2}.  It is perhaps possible that ED can also open a window into non-WIMP DM in the $\sim$ 
10 MeV to 1 GeV mass range which we will consider below.  What is immediately clear is that if EDs exist at the $\sim 10 -1000$ MeV scale then, 
to avoid conflict with many experimental constraints, none of the SM fields can experience these EDs and thus the SM must be confined to a 4-D brane within this larger 
bulk space.{\footnote {Interestingly, EDs of this (inverse) mass scale have arisen previously in discussions of the ADD model\cite {ED} when the number of 
EDs is chosen to be 6 or 7 and the low-energy Planck scale is not far above the current limits from the LHC $\sim 5-10$ TeV.}} Here we imagine that only the 
DM and the mediator field  for the DM interactions with the brane-localized SM particles are allowed to exist in this higher dimensional bulk, \ie, the 
{\it EDs are Dark}. For simplicity we will consider here the case of a single, flat, extra dimension (although generalizations of this framework can easily be 
constructed\cite{morrissey,keith}) in the spirit of Universal Extra Dimensions (UED)\cite {ED} while simultaneously ignoring the effects of gravity in the current discussion of this setup. 
However, it is interesting to speculate on the embedding of the approach that we will discuss here into a more complex structure that {\it does} include gravity and addresses the 
hierarchy problem; here the ADD model\cite{ED} is the natural choice as the ED in ADD are flat. Within this scenario, if the number of additional dimensions $n=5(6)$ and 
the ED reduced Planck scale is taken to be $\sim 10$ TeV, just above the current LHC constraints, then the ED would have an inverse size of roughly $R^{-1} \sim 100(1000)$ MeV 
which, as we'll see,  is comparable to that of interest to us here. (This result assumes, of course, that all these ED have the same size.) Thus it may be possible to embed the model 
setups that follows into an ADD-like scenario which may lead to some very additional interesting phenomenology. This is, however, beyond the scope of the present work.

One self-consistent and phenomenologically interesting scenario for DM at these mass scales is the Dark Photon (DP) model\cite{vectorportal} wherein a new 'dark' $U(1)_D$ 
gauge field, acting as the DM mediator, kinetically mixes\cite{KM} (\ie, via the `kinetic mixing portal') with the SM $U(1)_Y$ hypercharge gauge field at the 
renormalizable level. Such a mixing can be generated, \eg,  by loops of fields having both types of gauge charges. In such a setup, the SM fields will only interact 
with DM, which is an SM singlet but carries a $U(1)_D$ `dark' charge, via this kinetic mixing. In more realistic 4-D versions of these models, the DP and DM masses are 
generally uncorrelated, independent parameters. For example, while the mass of the DP is usually generated through spontaneous symmetry breaking, \ie, via the 
vacuum expectation value (vev) of a dark SM singlet Higgs field which also carries a $U(1)_D$ charge, the DM field, also being a SM singlet, can have a $U(1)_D$-invariant 
mass term whose value is generally unrelated to the dark Higgs vev. Why these two mass scales should be similar, as they must be to satisfy several 
phenomenological requirements, \eg, the value of the DM relic density, is somewhat of a mystery. Furthermore, in such models, there is no way to avoid some coupling of the dark 
scalar with the SM Higgs through a renormalizable term. This leads to an additional Higgs portal interactions\cite{higgsportal} between the DM and the SM of 
some significance making the resulting physics more complex and can lead to too large of a branching fraction for exotic SM Higgs decays 
unless the quartic coupling linking the dark and SM Higgs sectors is tuned to a very small value. Fortunately, as we will see, we can employ ED boundary conditions to completely 
turn off this Higgs portal coupling. In this paper we will also see that the similarity of the DP and DM masses can be a natural outcome of a scenario 
with ED. Furthermore, the same choice of boundary conditions in the ED can be employed to break the $U(1)_D$ symmetry and generate 
a DP mass without the need to introduce the additional dark Higgs field, though such fields may find other applications. In order to satisfy important CMB 
constraints on the DM annihilation cross section for masses in our range of interest, we consider the DM to originate from a complex bulk scalar field that may or may 
not obtain a vev. This choice naturally splits this general setup into two distinct model classes with very different phenomenologies that we will separately explore in detail. The 
specific pair of Abelian model classes that we present here are to be thought of as only toy models-proofs of principle and are not fully detailed, UV-complete scenarios.
 
The outline of this paper is as follows: In Section 2 we provide an overview of the essential ingredients of the $U(1)_D$ gauge sector and some of the important model building 
constraints that need to be addressed in our ED constructions. This includes an analysis of the 5-D kinetic and mass mixing between the 5-D DP Kaluza-Klein (KK) tower 
states with those of the SM where we demonstrate the need for a brane-localized kinetic term (BLKT)\cite{blkts} required to bring the effective 4-D action to the usual canonical 
form while avoiding tachyons and ghosts in the spectrum. The constraints imposed by measurements of the CMB on the nature of the DM field and the relative 
DM/DP mass spectra are also presented in this Section. In Section 3 we provide detailed discussions of a pair of model classes wherein the DM is a assumed to be a complex scalar 
which does or does not obtain a vacuum expectation value. The phenomenology of these two model classes is discussed and shown to be quite distinctive and leading 
to interesting signatures in future experiments. A discussion and our conclusions can be found in Section 4.



\section {Essentials of the DP Framework in One Extra Dimension}
\label{section:Model}

This Section provides an overview of the essential elements of the ED models that we consider below as well as some general model-building requirements on such theories.

\subsection{Kinetic and Mass Mixing of the KK Dark Photon}

We begin our analysis by considering a straightforward generalization of the usual DP portal/kinetic-mixing (KM) models to the case of one flat ED. Here we write the SM  
plus pure $U(1)_D$ gauge field parts of the full 5-D action as 
\begin{equation}
S=\int ~d^4x ~\int_{y_1}^{y_2} ~dy ~\Big[-\frac{1}{4} \hat V_{AB} \hat V^{AB}  ~+\Big(-\frac{1}{4} \hat B_{\mu\nu} \hat B^{\mu\nu} 
+\frac{\epsilon_5}{2c_w} \hat V_{\mu\nu} \hat B^{\mu\nu}  + L_{SM} \Big) ~\delta(y-y_{SM}) \Big] \,,  
\end{equation}
where we consider the extra dimensional co-ordinate, $y$, to take on values on an interval bounded by two branes at $y=y_{1,2}$, \ie, $y_1  \leq y \leq  y_2$; the SM is assumed to 
be confined to {\it one} of these two branes, \ie, $y_{SM}$. Here $L_{SM}$  is the remainder of SM Lagrangian, apart from the hypercharge field strength piece, written explicitly 
above, $\hat V$ is the 5-D DP gauge field while $\epsilon_5$ is a 5-D kinetic mixing parameter, normalized in a familiar manner\cite{stuff} with $c_w=\cos \theta_w$.  Note 
that since all the SM fields are localized to one of the branes, the kinetic mixing itself must also be localized there. Following the usual procedure, we expand the the 5-D DP gauge 
field into a Kaluza-Klein (KK) tower of states, separating the 4-D and $5^{th}$ components: 
\begin{equation}
\hat V^{\mu [5]}(x,y) = \sum_n ~ f_n^{[5]}(y) \hat V_n^{\mu [5]} (x)\, 
\end{equation}
where the functions $f_n^{[5]}(y)$ are determined by the equations of motion and the specific boundary conditions (BCs) as usual. (For 
convenience we will work in the $V^5=0$ gauge in this Section.) We remind the reader that in performing this very familiar procedure, an integration by parts 
is necessarily performed and that the BC  $f_m \partial_y f_n |_{y_1}^{y_2} =0$ 
for all values of $n,m$, is imposed. Here we  employ the common shorthand notation for the difference of the values at either end of the interval. In what follows 
we will consider specific BCs which will satisfy this requirement; recall that in the most typical discussed scenario employing orbifold BCs that this requirement is trivially satisfied. 
Integration over the ED then produces the standard results except for some distinct aspects, the most important being that the single 5-D KM term above now becomes 
an infinite sum of KM terms in 4-D given by
\begin{equation}
\sum_n \frac{\epsilon_n}{2 c_w} \hat V_n^{\mu\nu} \hat B_{\mu\nu}\, 
\end{equation}
where we identify $\epsilon_n = \epsilon_5 f_n(y_{SM})$, \ie, in principle each member of the DP KK tower can experience a different amount of KM depending upon the value of its wavefunction 
on the SM brane. As we will see, consistency of the field redefinitions introduced to remove the KM {\it requires} that this be so.  As usual, we can think of the non-zero $\epsilon$'s as 
arising from, \eg, the existence of a pair of SM color- and iso-singlet fields with slightly different masses localized on the SM brane and having hypercharge(dark charge) $Y(Q_D)=1(1)$ 
and $1(-1)$, respectively\cite{KM}. 

At this point in the standard 4-D treatment, we would `undo' the KM via a set of field redefinitions to bring the action into canonical form; we must do the same here but now face an infinite KK 
tower of states. This `undoing' naturally leads to a BLKT for the DP field; the simplest way to see this is the following. Suppressing Lorentz indices and performing a shift 
$\hat B\to B+\frac{\epsilon_5}{c_w} \hat V$ on the the SM brane removes the KM but leaves a term $\frac{\epsilon_5^2}{4c_w^2}~\hat V^2$ which corresponds to a {\it negative} 
(dimensionless) BLKT $\delta_0=-\frac{\epsilon_5^2}{Rc_w^2}$. A negative BLKT is well-know to lead to tachyonic and/or ghost states in the KK mass spectrum 
so that this negative BLKT must be necessarily compensated by already having in the setup an additional {\it positive} BLKT on the SM brane. Another way to see this makes direct use of the KK 
decomposition and generalizes the treatment of the KM of two new $U(1)$'s with the SM hypercharge\cite{Heeck:2011md}.  Writing $\hat B = B+\sum_n \alpha_n V_n$ and defining $\Sigma_i= 
(1- \sum_{a=1}^i \epsilon_a^2/c_w^2)^{1/2}$, some algebra informs us that 
\begin{equation}
\alpha_n=\frac{\epsilon_n}{c_w} ~\frac{1}{\Sigma_n\Sigma_{n-1}}\,, 
\end{equation}
and we now suggestively re-write
\begin{equation}
\Sigma_n^2= 1-\frac{\epsilon_1^2}{c_w^2} \sum_{a=1}^n \frac{\epsilon_a^2}{\epsilon_1^2}\,.
\end{equation}
Note as we ascend further up the KK tower the sums appearing in the $\Sigma$'s extend to larger and larger values of $n$, and eventually to infinity. Clearly if all the $\epsilon$'s had the 
same value or were to grow in magnitude with increasing $n$, then at some point the $\Sigma$'s would become {\it undefined}. However, a BLKT on the SM brane for 
the DP field leads to the well-known effect of suppressing the values of the the KK wavefunctions ever more strongly with increasing $n$ and this leads to a convergence in the 
sums encountered above provided that the BLKT is of sufficiently magnitude.  We note that once this convergence is assured, for typical values of $\epsilon_1 \leq 10^{-(3-4)}$, we can 
work to leading order in the $\epsilon_n$ and employ the approximate result that $\Sigma_n \simeq 1$ in our subsequent numerical calculations. In that case, to leading order in the 
$\epsilon_n$, we find  that we can replace $\hat V \to V$\cite{Heeck:2011md} as non-trivial terms only enter here at second order in these parameters.

\subsection{Mixing Phenomena}

 Once we go to the canonically normalized basis for the hypercharge and DP KK fields we next consider the mass mixing that results between the SM $Z$ and this DP KK tower whose 
members now couple weakly to hypercharge (and thus to the SM Higgs) via the $\epsilon_n \neq 0$. Similarly, since the $Z$ mixes with all of the $V_n$ KK tower members it also picks up a 
small coupling to all of the dark sector fields. This mixing leads to a number of interesting effects even in the 4-D case. We note that at this point, to be quite general, we 
will not commit ourselves with respect to any specific ED origin of the {\it unmixed}  DP KK mass terms or their specific 
values other than to note that they will generally be assumed to be of order of the inverse size of the compactification scale, $R^{-1} \sim 10-1000$ MeV. We will denote these KK masses as 
$M_n$ which is sufficiently general for present purposes. We can schematically write the resulting neutral gauge boson mass-squared matrix, ${\cal M}^2$, in terms of $M_Z^2$, 
the set of $\epsilon_n$'s and the corresponding unmixed DP KK squared masses,  $M_n^2$, obtaining  the general form (with each element to leading order in the $\epsilon_n$'s)
\begin{eqnarray}
{\cal M}^2 & =  \left( \begin{array}{cccc}
                         M_Z^2 & -t_w\epsilon_1 M_Z^2 & -t_w \epsilon_2 M_Z^2   & ... \\
                          -t_w \epsilon_1 M_Z^2 &  M_1^2+t_w^2 \epsilon_1^2 M_Z^2 & t_w^2 \epsilon_1 \epsilon_2 M_Z^2 & ....\\
                          -t_w \epsilon_2 M_Z^2 & t_w^2 \epsilon_1 \epsilon_2 M_Z^2 & M_2^2+ t_w^2 \epsilon_2^2 M_Z^2  & ....\\
                          ... & ... & ... & ...
                         \end{array}\right) \,.
\end{eqnarray}
where $t_w =\tan \theta_w$ and reproduces the 4-D result to this order. (See, \eg, the last paper in Ref.~\cite{vectorportal}.)  To leading order in the $\epsilon$'s this matrix can 
be diagonalized by the small rotations:
\begin{eqnarray}
V_i &\to& V_i +t_w \frac{\epsilon_i M_Z^2}{M_i^2-M_Z^2}~Z\nonumber\\
Z &\to& Z -t_w \sum_i \frac{\epsilon_i M_i^2}{M_i^2-M_Z^2}~ V_i \,,
\end{eqnarray}
which lead to the corresponding mass shifts for the $Z$ and DP KK states given by\cite{vectorportal}
\begin{eqnarray}
 M_i^2 &\to& M_i^2\Bigg[1 + \frac{t_w^2\epsilon_i^2 M_Z^2}{M_i^2-M_Z^2} \Bigg]\nonumber\\ 
 M_Z^2  &\to&  M_Z^2~\Bigg[ 1- t_w^2M_Z^2 \sum_n ~\frac{\epsilon_n^2}{M_n^2-M_Z^2}\Bigg]\,
\end{eqnarray}
We note the appearance of {\it infinite} KK sums in the above expressions, \eg,  for the $Z$ mixing and mass shift. As will be seen below, the presence of the BLKT will allow these sums 
to converge due to the rapidly decreasing values of the $\epsilon_n^2$'s. Even with such convergence one may worry that the shift in the $Z$ mass will be small enough as to not be in 
conflict with the agreement of the SM with the current electroweak data\cite{pdg}.  Another concern is the possibility that  $Z-$KK mixing may significantly alter the various $Z$ couplings 
to the SM fields and, since the $V_n$ couple to the dark sector,  lead to a sizable contribution to the $Z \to $ invisible decay width (and possibly for 
the SM Higgs field as well). We will see below that these issues are all under control within specific model frameworks. 
Whatever the detailed nature of the DM sector (scalars or fermions), the $V_i$ will couple to pairs of DM tower states, $\rm DM_n \rm DM_m$, with a strength which we will denote as 
$g_D \tilde c_{nm}^i$ below where here $g_D$ is the 4-D dark gauge coupling and are given by integrals over $y$ co-ordinate of the product of the relevant gauge and DM KK tower 
wavefunctions.  The mixing of the $Z$ with the DP KK modes also induces a corresponding coupling of the $Z$ to pairs of DM states, $\rm DM_n \rm DM_m$, which is given by
\begin{equation}
-g_D \epsilon_1 t_w \sum_i \frac{(\epsilon_i/\epsilon_1) \tilde c_{nm}^i M_Z^2}{M_i^2-M_Z^2}\,.
 \end{equation}
If the $\epsilon$'s and $\tilde c$'s were independent of their indices this sum would not be well-behaved; BLKTs, as we will see, insure convergence.
We note that for numerical  purposes $g_D$ will here always be taken to be {\it not far} from $O(0.1-1)$ (like $e$ or the SM weak charge) so that $g_D^2>>(e\epsilon_i)^2$. 
Since the unmixed $V_i$ couple to hypercharge, their mass mixing with the $Z$ alters their couplings as well as those of the $Z$. To leading order in the $\epsilon$'s this mixing  
produces $V_i$ couplings to the SM fields of the form 
 \begin{equation}
 \frac{g}{c_w}  t_w \epsilon_i \Bigg[ T_{3L} \frac{M_i^2}{M_Z^2-M_i^2} + Q \frac{c_w^2 M_Z^2-M_i^2}{M_Z^2-M_i^2}\Bigg]\,,
 \end{equation}
 where $Q(T_{3L})$ is the SM electric charge (third component of weak isospin). For the lighter members of the KK tower with masses below $\sim$ a few GeV, where the ratios 
 $M_i^2/M_Z^2$ can be safely neglected, the $V_i$ couple to the combination $\simeq eQ\epsilon_i$ since $e=gs_w$. Thus lighter members of the $U(1)_D$ 
 tower will all couple to the SM like the usual 4-D dark photons but with a decreasing strength with increasing $i$ since the $\epsilon$'s must decrease as $i$ increases.  However, in the opposite 
 limit where the ratios $M_Z^2/M_i^2$ are now small, the $V_i$ couple to $\simeq \frac{g}{c_w} t_w\epsilon_i Y$ where $Y=Q-T_{3L}$ is the SM hypercharge. Note that as we ascend the 
 gauge KK tower, once $M_i^2/M_Z$ becomes significant the tower field picks up a parity-violating interaction with the SM fields which can potentially lead to additional constraints. 
 
 In the case of the $Z$, to lowest order in the mixing, employing the tree-level relationship $e=gs_w$ as above, the $Z$ couplings including mixing effects can be written as
 \begin{equation}
 \frac{e}{s_wc_w} \Big[(1+F)T_{3L}-(s_w^2+F)Q\Big]\,,
 \end{equation}
 where we have defined the dimensionless quantity, $F$, as   
 \begin{equation}
 F= \sum_i \frac{(t_w\epsilon_i)^2 M_Z^2}{M_Z^2-M_i^2} \,,
 \end{equation}
 which is related to the $Z$ mass shift obtained above: $\delta M_Z^2/M_Z^2 =1-F$.  Defining $\alpha_w=\alpha_{QED}(M_Z^2)$ we can use this expression and that 
 for the (null) shift in the $W$ boson mass to determine the (tree-level) values of the STU oblique parameters\cite{peskin,maksymyk}  to lowest order in the parameter $F$:
 \begin{eqnarray}
 T&=&\frac{2F}{\alpha_w}\nonumber\\
 S&=&\frac{4c_w^2F}{\alpha_w}\nonumber\\
 U&=&0\,,
 \end{eqnarray} 
 where we see that $S$ and $T$ are related as $S=2c_w^2T$. With values of $\epsilon_1$ below $\sim 10^{-(3-4)}$ we will have no tree-level conflict with the usual electroweak data. 
 Lastly, in a similar manner, the mixing of $Z$ with the gauge KK states (the off-diagonal terms in the mass matrix) induces a set of HZ$V_i$ couplings, where $H$ denotes the SM 
 Higgs field, and which are given to LO in this mixing by
 \begin{equation}
 K_{HZV_i}=\frac{2M_Z^2}{v_H}~\Bigg[\frac{t_w\epsilon_i M_i^2}{M_Z^2-M_i^2}\Bigg] \,,
 \end{equation}
 while the corresponding $HV_iV_j$ coupling is given by
 \begin{equation}
 K_{HV_iV_j}\simeq \frac{2M_Z^2}{v_H}~\Bigg[\frac{t_w\epsilon_i M_i^2}{M_Z^2-M_i^2}\Bigg]~\Bigg[i \to j\Bigg] \,,
 \end{equation}
where in both cases we have normalized to the SM $HZZ$ coupling; here $v_H$ is the usual SM Higgs vev $\sim 246$ GeV. These expressions show that these couplings have an unusual 
dependence on the location of the relevant $V_i$ within the gauge KK tower in that $M_i^2$ (clearly) grows with $i$ while the $\epsilon_i$'s will decreases with increasing $i$.

\subsection{Cosmology: Planck and CMB Constraints}

In the 4-D DP scenario, it is well known that the measurements of the CMB power spectrum by Planck\cite{Ade:2015xua} place constraints on the nature of the DM and the relative 
DM/DP mass spectrum when these states are light, \ie,  below a few GeV in mass. This remains true in the 5-D extensions of these models that we consider. These 
constraints essentially place bounds on the DM annihilation cross section into various electromagnetically interacting final states that can lead to re-ionization in the early universe with the 
sensitivity peaking near $z \sim 600$\cite{Liu:2016cnk}.  For example, for 50-100 MeV DM that has an s-wave annihilation into the $e^+e^-$ final state (which we might expect to be the dominant 
mode in this mass range), the annihilation cross-section is constrained to roughly satisfy the bound  $<\sigma v> \lsim  (1-3)\cdot 10^{-29} ~\rm cm^3 \rm s^{-1}$ 
This is $\sim$ 3 orders of magnitude smaller than the canonical thermal cross section needed at freeze-out to achieve the observed relic density\cite{Steigman:2015hda},
\ie,   $<\sigma v>\simeq 4.5\cdot 10^{-26} ~\rm cm^3 \rm s^{-1}$ for (self-conjugate) DM in this mass range.  This apparent 
conflict is only a serious one if the value of the DM annihilation cross section, $<\sigma v>$, does not vary significantly with temperature or, more precisely, with velocity. At $z \sim 600$ the 
DM is moving far slower than it would be at freeze-out which occurs at far higher temperatures so that a significant velocity-dependence in the annihilation rate can significantly soften or even 
eliminate this as a serious constraint. A straightforward solution to this problem is to consider only those models wherein the DM annihilation to SM particles via the spin-1 DP 
tower is a {\it p-wave} process so that $<\sigma v>$ has an overall $v^2$ suppression factor. In 4-D, following this approach 
excludes the possibility of Dirac fermionic DM as these lead to an s-wave annihilation process without some additional physics. However, the DM can still be, \eg,  a complex scalar or a 
Majorana fermion and the possibility of co-annhilation opens up further possibilities. 5-D models are similarly constrained by these same considerations.  Furthermore, one finds in the 4-D case 
that an additional problem arises if  $M_{DM}  > M_{DP}$. If 
this occurs the DM can annihilate into pairs of DPs which is found to be an s-wave process for both fermionic and bosonic DM in the initial state\cite{Arcadi:2017kky} and is thus constrained 
by the CMB. In the 5-D case, this tells us that the lightest mode in the DM KK tower (\ie, the actual DM) must be lighter than the corresponding lightest KK state in the 
DP mediator tower which can be a significant model-building constraint. This is particularly true in the case of the 5-D models we consider below as both of these 
masses will have a related ED origin of order $R^{-1}$.  Note that here we are particularly interested in scenarios wherein the lightest DP KK and DM masses are 
naturally similar in value as this produces the most successful phenomenology particularly with respect to the relic density. Our goal is to avoid putting this mass relationship into the model 
`by hand' as is generally done in 4-D where these masses can {\it a priori} be vastly different. We note that the choice of p-wave annihilation implies, apart from some unusual circumstances, that the 
indirect detection of DM annihilation {\it today} will be very unlikely due to the small value of the present-day annihilation cross section arising from this strong $v^2$ suppression. 

In principle, there may be additional constraints on a set of short-lived DP KK states in the the $\sim 0.1-1$ GeV mass range arising from other cosmological considerations. In 4-D there has been 
only a limited examination of the impact of a single DP for such masses\cite{pair} where it has been shown that most of the sensitivity is at very small values of $\epsilon$ below our range of 
interest. This subject is certainly deserving of further study.

\section{Models With Complex Scalar DM}

Since scalar DM allows for a simple means to obtain a p-wave annihilation cross section for DM we will concentrate on this possibility and entertain other possibilities elsewhere\cite{wip}.  This 
scalar, $S$, is naturally a complex field as it must carry a dark change for it to couple to the DP. Within this general framework there are two possible scenarios which depend upon whether 
or not this complex scalar obtains a vev. In 4-D such a vev is required in order to break the $U(1)_D$ symmetry and give a mass to the DP. In 5-D, however, this need not be the case as we can break 
this gauge symmetry by the appropriate choice of BCs as will be seen in detail below. The phenomenology of these two possible paths is quite different and we consider them separately in the 
following. In either case one may be concerned that $S$ (or its real part $h$) will mix with the SM Higgs through a generalization of the (brane localized) quartic term
\begin{equation}
 S_{HS} =\int ~d^4x ~\int_{y_1}^{y_2} ~dy  ~\lambda_{HS} H^\dagger H S^\dagger S~\delta(y-y_{SM}) \,
\end{equation}
and possibly lead to significant Higgs physics alterations without fine-tuning the coupling. In our ED scenarios, as the SM Higgs would then couple with the entire KK tower of dark Higgs 
states this would likely lead to unacceptable changes in the SM Higgs properties and decay modes.  Fortunately in a 5-D setup we can avoid this coupling completely by requiring the appropriate 
BCs so that  the KK tower of states with which the Higgs would mix have vanishing values for their 5-D wavefunctions on the SM brane. The implementation of this condition will 
differ in the two scenarios we consider below.

\subsection{Model 1: Complex Scalar DM Without a VEV}

The simplest possible 5-D DM matter within the KM scenario previously described is to add a complex 5-D scalar, which is a SM singlet field, to the action above. Here we 
assume that its bulk potential is such that it does not produce a 5-D vev and for simplicity we set its bulk mass term to zero as it will play no essential role in what follows. Ordinarily such mass terms 
for scalars would be of significant importance as one normally applies orbifold BCs, \ie, either $S$ or $\partial_y S$ vanishes simultaneously on both branes, so that a massless zero-mode is 
present with the bulk mass being the only source of mass for the lightest KK scalar mode. This will not be the case here.

The first step in is to perform a KK decomposition of the 5-D part of the action; this is given by the expression 
\begin{equation}
S_{5D}=\int ~d^4x ~\int_{y_1}^{y_2} ~dy ~\Big[-\frac{1}{4}  V_{AB}  V^{AB}  + (D_A S)^\dagger (D^A S) -V(S^\dagger S) \Big]\,
\end{equation}
where $D_A$ is the 5-D covariant derivative, $D_A=\partial_A +ig_{5D} Q_D V_A$, and where $Q_D$ is the dark charge of $S$ with $g_{5D}$ being the $U(1)_D$ coupling. 
Note that we will drop consideration of any of the potential terms in this subsection appearing in $V(S^\dagger S)$. The next step is to KK decompose both $S$ and $V$; for the moment 
let us ignore the possible existence of any BLKTs. In that case both $S$ and $V$ can be expanded in a similar manner. In thus setup it is convenient to adopt the unitary, 
$V_5=0$, gauge in our analysis below.  We further assume that the boundaries at $y_1=0$ and $y_2=\pi R$ define the interval of interest and that for model building purposes 
we place the SM fields at $y_{SM}=\pi R$.

Generally we can write the forms of the $V,S$  5-D KK wavefunctions for these fields in suggestive notation as 
\begin{eqnarray}
v_n(y)&=&A \cos \kappa y + B \sin \kappa y\nonumber\\
s_n(y)&=&C \cos \sigma y + D \sin \sigma y \,, 
\end{eqnarray}
with all of the parameters $A-D,\kappa,\sigma$ generally being $n$-dependent, and determined by both the BCs and the normalization conditions.  What are our BCs? 
We first require that $s_n(\pi R)=0$ (in order to avoid the Higgs portal) while the $v_n$ cannot vanish there or the corresponding fields would fail as mediators. Remembering the integration 
by-parts constraint above {\it and} that we wish to avoid the possibility of massless modes for any of these KK states we choose the following BCs: $\partial_y v_n(\pi R)=v_n(0)=0$ while 
$\partial_y s_n(0)=s_n(\pi R)=0$. Note that if we had chosen the usual orbifold BCs for the vectors a massless zero-mode would 
appear. Here, the (degenerate) masses of both KK towers of states are given by $m_n=(n+\frac{1}{2})/R$, where $n=0,1,..$, and where 
$v_n \sim \sin x_ny/R$ and $s_n \sim \cos x_ny/R$  with $x_n=n+1/2$. {\footnote {Since the lowest mass state, $V_1$, corresponds to the choice $n=0$ with this counting, the 
index $n=0$ will sometimes by applied as a lowest mode label.}}  
We observe that there are no massless modes and that we have broken the gauge symmetry without an explicit Higgs vev. 
Of course in the $V_5=0$ gauge employed here, the $V_5$ field has been eaten level-by-level to become the longitudinal Goldstone component of {\it all} of the massive vector KK tower fields. 

\begin{figure}[htbp]
\centerline{\includegraphics[width=5.0in,angle=90]{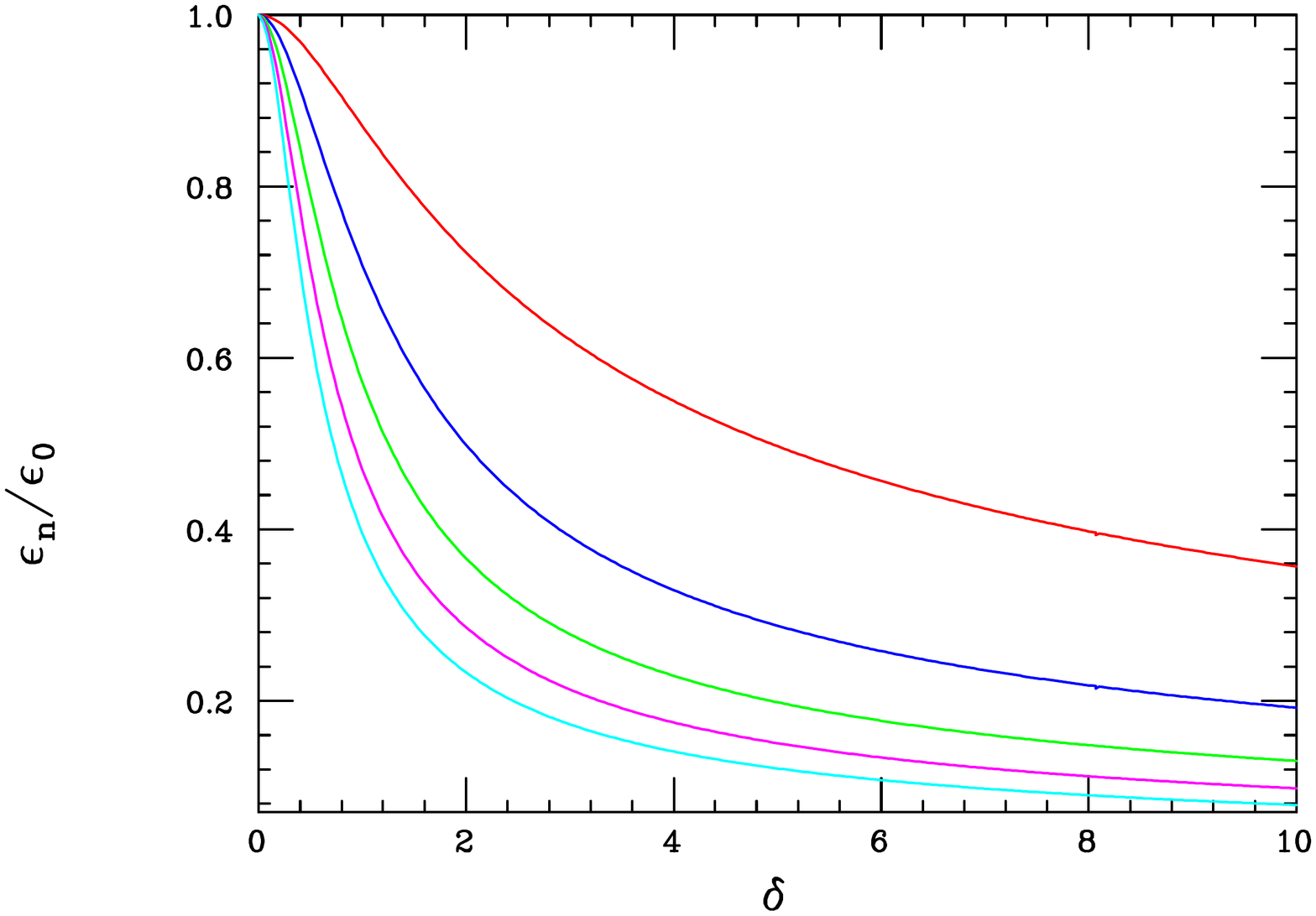}}
\vspace*{-3.1cm}
\centerline{\includegraphics[width=5.0in,angle=90]{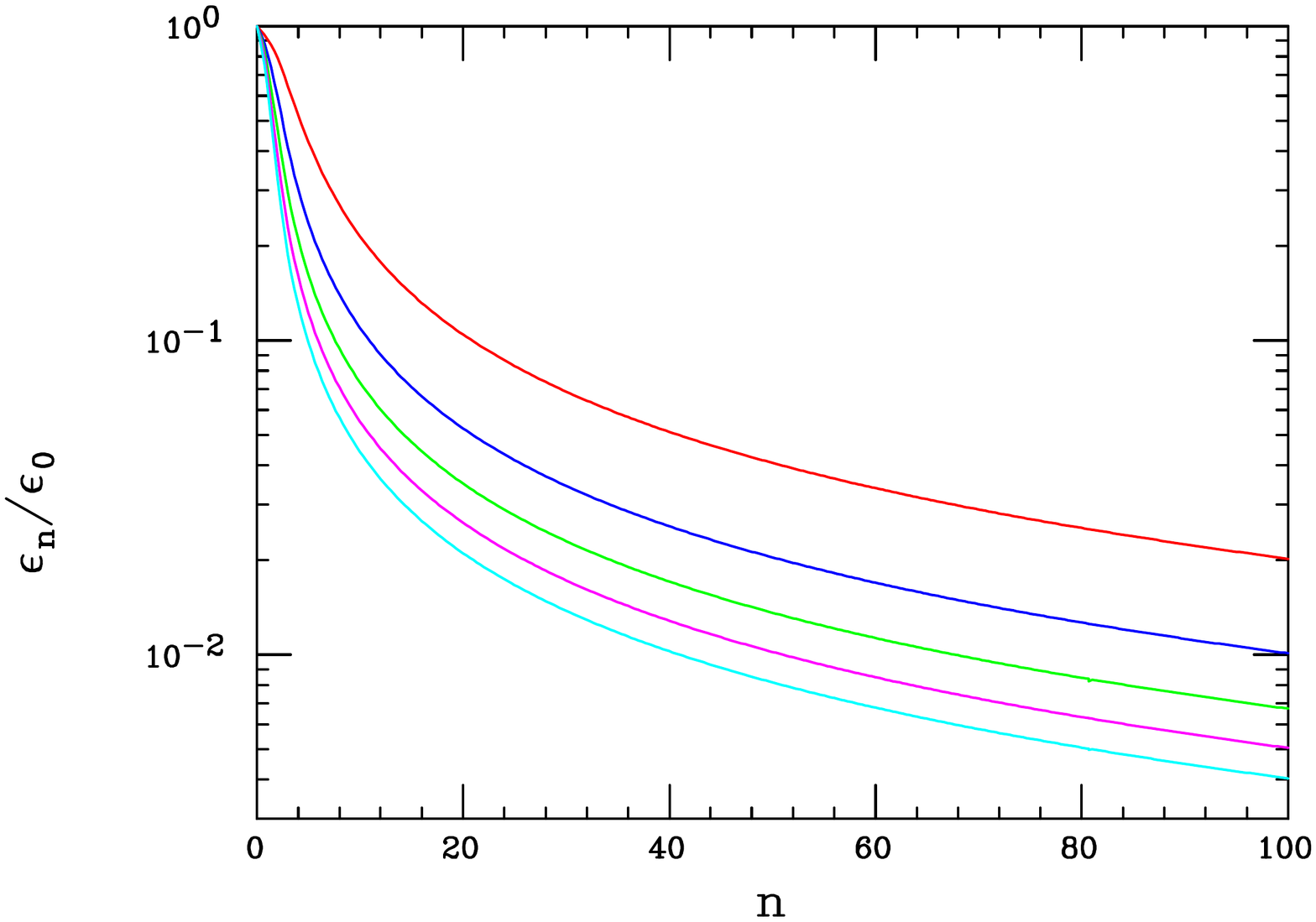}}
\vspace*{-1.90cm}
\caption{Top:$\epsilon_n/\epsilon_{n=0}$ as a function of $\delta_A$ for, from top to bottom, $n=1,2,..5$. Bottom: the same ratio but now as a function of $n$; from top to 
bottom the curves are for $\delta_A=1,2,..5$.}
\label{eps1}
\end{figure}

\begin{figure}[htbp]
\centerline{\includegraphics[width=5.0in,angle=90]{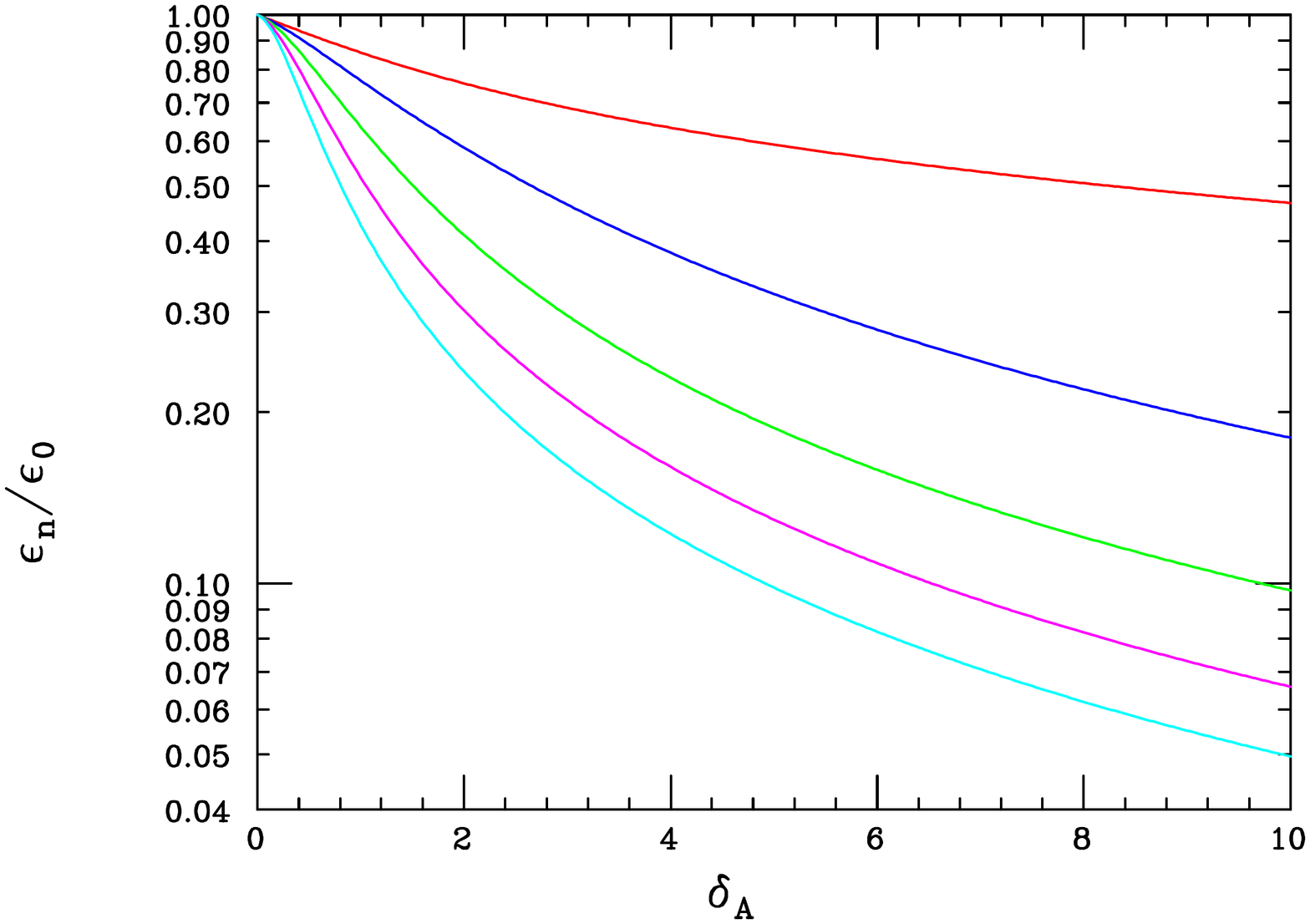}}
\vspace*{-3.1cm}
\centerline{\includegraphics[width=5.0in,angle=90]{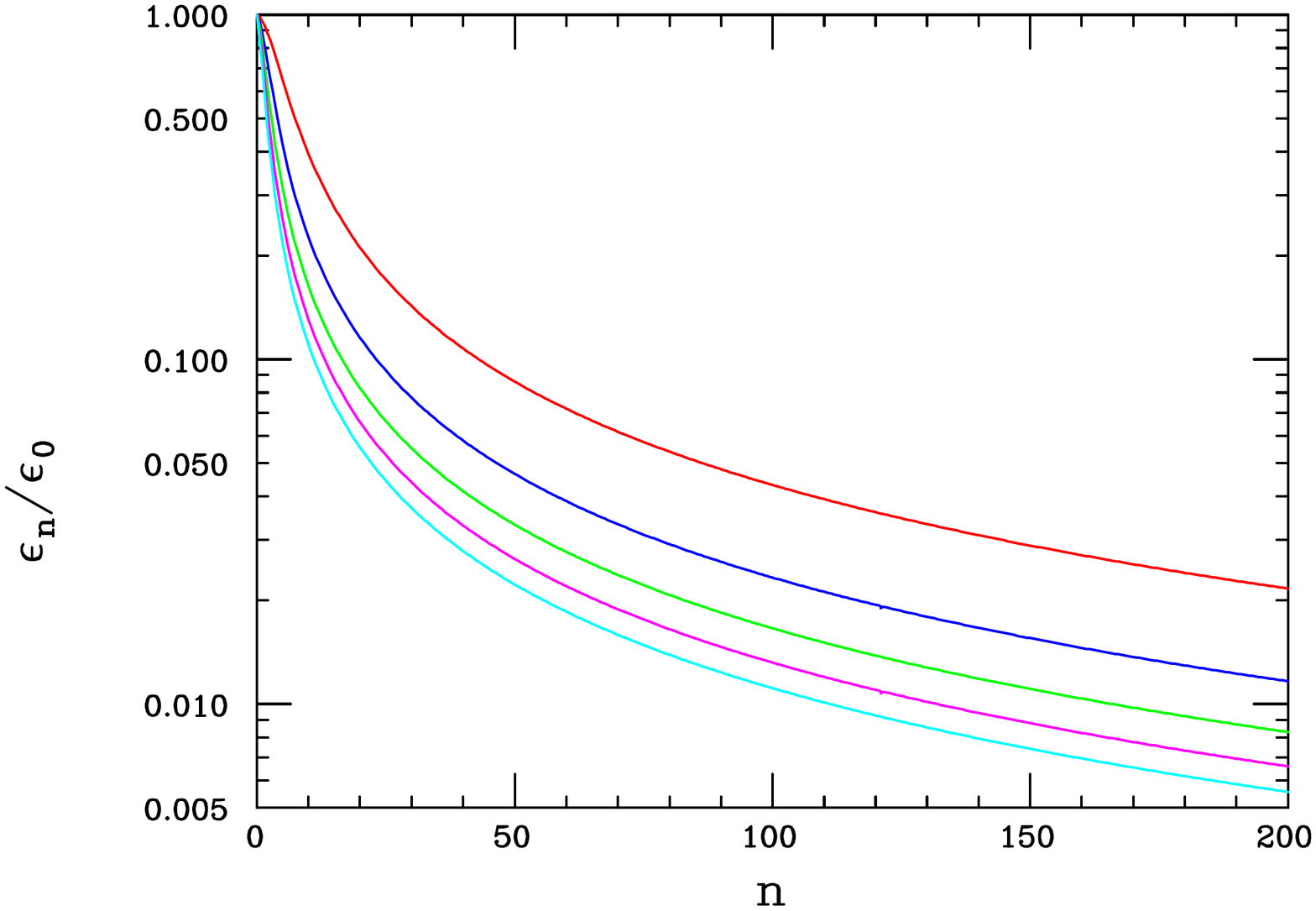}}
\vspace*{-1.90cm}
\caption{Top: Same as the top panel in the previous Figure but now shown on a log scale.Bottom: the sane as the lower panel in the previous Figure but now emphasizing smaller 
values of $\delta_A$, from top to bottom, the curves correspond to $\delta_A=0.5,1,..,2.5$.}
\label{eps2}
\end{figure}

\begin{figure}[htbp]
\centerline{\includegraphics[width=5.0in,angle=90]{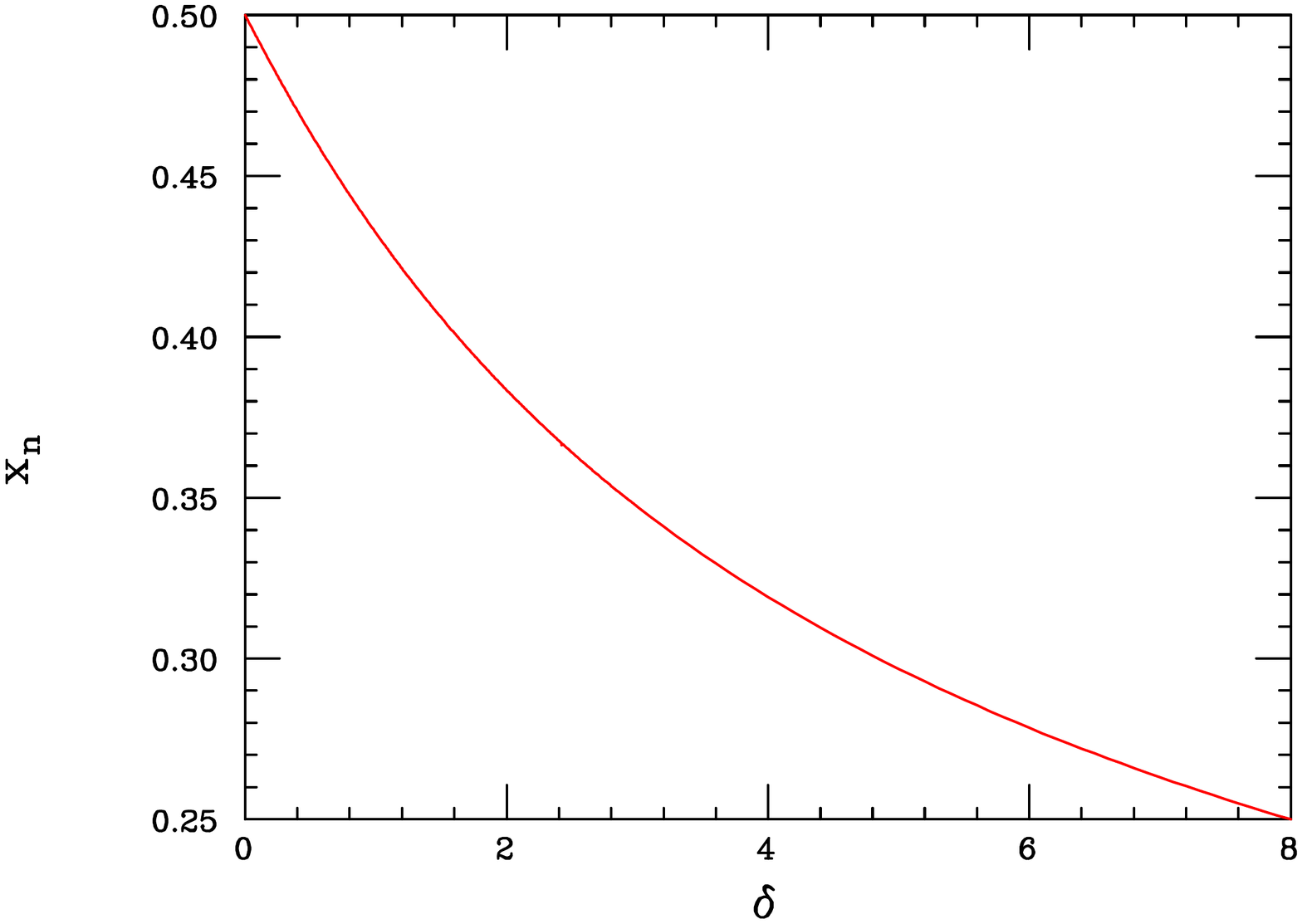}}
\vspace*{-3.1cm}
\centerline{\includegraphics[width=5.0in,angle=90]{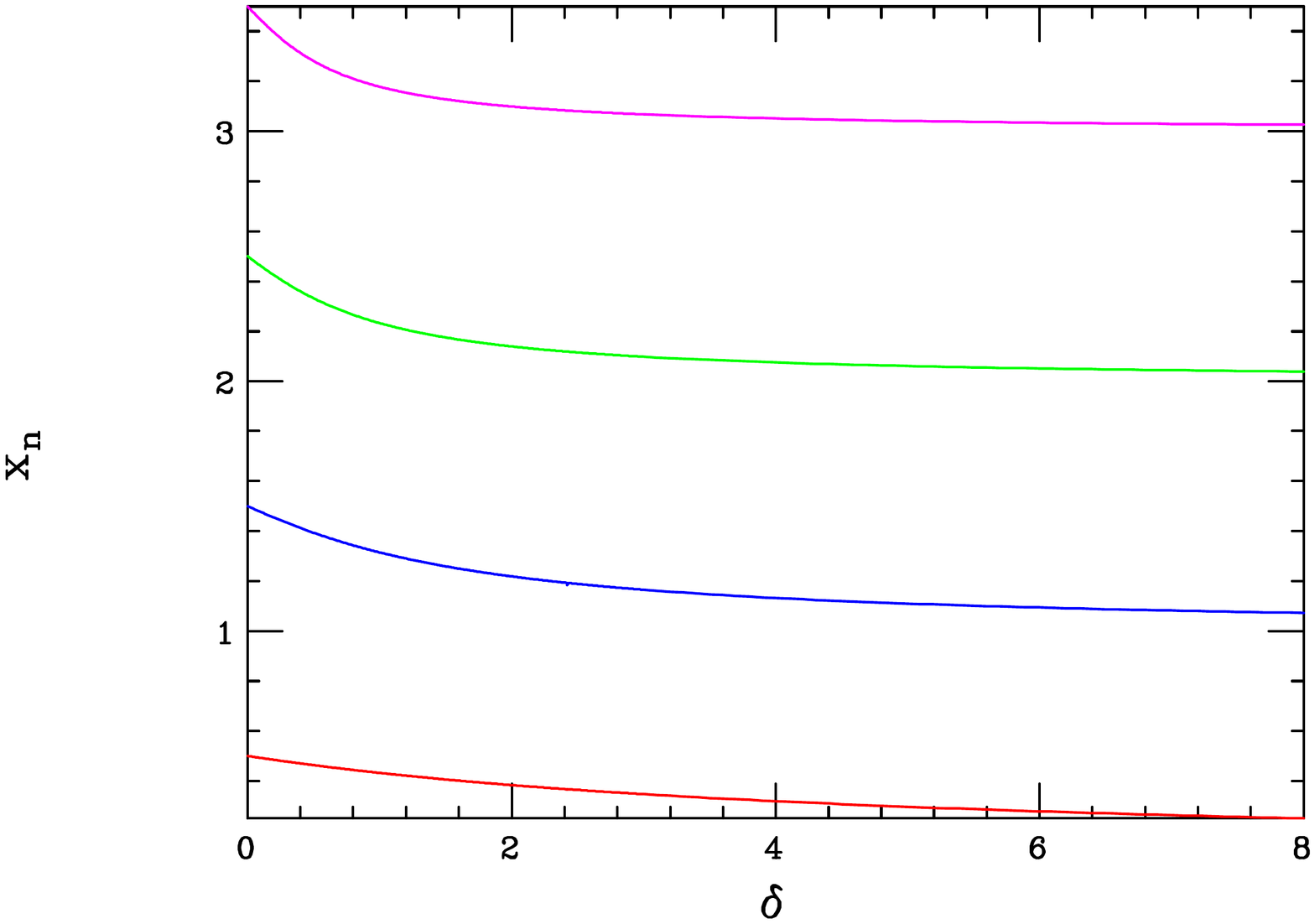}}
\vspace*{-1.90cm}
\caption{Top: Lowest lying root $x_{n=0}$ as a function of the BLKT parameter $\delta$ (for both vectors or scalars) as we discuss in the text. Bottom: Same as top panel but 
now, from bottom to top, for $x_{n=0,1,2,3}$.}
\label{masses}
\end{figure}

While this appears promising, we've not been entirely successful since at this point all of the $\epsilon_n$'s will have the same value {\it and} the two sets of gauge and scalar KK states are mass 
degenerate while from our discussion above we require that $m_{S_1} < m_{V_1}$ to avoid there being an $s$-wave DM annihilation. Since these two effects are directly related 
to the values of the $v_n(\pi R)$, in order to reduce these quantities as $n$ increases, we introduce the anticipated BLKT,  whose coefficient we denote by a dimensionless 
parameter $\delta_A$,  for the $v_n$ on the SM brane. (The presence of a corresponding BLKT on the $y=0$ brane would have no effect as the $v_n$ vanish there.) This adds a term
\begin{equation}
S_{BLKT}=\int ~d^4x \int_{0}^{\pi R} dy ~\Big[-\frac{1}{4}  V_{\mu\nu}  V^{\mu\nu}  \cdot\delta_AR~\delta(y-\pi R) \Big]\,
\end{equation}
to the gauge part of the action.  We know from our previous discussion that such a term {\it must} be present {\it a priori} and, further, that $\delta_A\geq |\delta_0|$ encountered above. 
This BLKT\cite {blkts} will introduce a discontinuity in the derivative of the $v_n$ wavefunctions at $y=\pi R$ and then one finds that the gauge 
KK tower masses $m^V_n=x^V_n/R$ will now be determined by the roots of the equation 
\begin{equation}
\cot \pi x^V_n= \frac{\delta_A}{2} x^V_n\,
\end{equation}
and the $v_n$ for $y\leq \pi R$ are now given by $v_n(y)=N^V_n \sin \big(\frac{x^V_n y}{R}\big)$ where the wave-function normalization is given by 
\begin{equation}
(N^V_n)^2 = \frac{2}{\pi R}~\Bigg[1+\Big(\frac{\delta_A x^V_n}{2}\Big)^2+\frac{\delta_A}{2\pi}\Bigg]^{-1}~\Big({\rm sin^2} \pi x^V_n\Big)^{-1}\
\end{equation}
The values of the $\epsilon_n$ can be easily read off from this normalization factor; here we take the convention that the $\epsilon_n$'s will be defined to be positive which 
results in additional powers of -1 appearing in some of the expressions below. 
$\delta_A\neq 0$ is observed to have multiple effects: As required, as $n$ increases the values of the ratio $\epsilon_n/\epsilon_1$ 
decrease roughly as $\sim 1/n$ as is shown in Figs.~\ref{eps1} and ~\ref{eps2}.
In these Figures we see that for fixed $\delta_A$ the value of $\epsilon_n$ falls rapidly with $n$ and also that, for a given KK mode, $n$, the value of $\epsilon_n$ decreases as 
$\delta_A$  increases.  
There is, however a further effect; when $\delta_A>0$ the entire KK mass spectrum of the field is lowered as seen in Fig.~\ref{masses}. {\footnote {We remind the reader 
that  if {\it negative} values of $\delta_A$ are employed this leads to ghosts and/or 
tachyons\cite{blkts} in the KK mass spectrum so that this possibility is excluded.}} The specific decrease of the values of the roots $x^V_n$ are (for 
$\delta_A>0$)  shown in Fig.~\ref{masses} as functions of $\delta_A$. 
Here we see that the action of the BLKT for large $\delta_A$ for $n\geq 1$ is to effectively reduce the root spectrum values, $x_n$, from $\simeq (n+1/2) \to ~\simeq n$. 

Since the $\epsilon_n$'s can now be explicitly calculated, we can evaluate some of the quantities discussed earlier, \eg, the oblique 
parameters were shown to depend upon a parameter $F$, defined above. Taking, \eg, $\delta_A=0.5$, $R^{-1}=100$ MeV and $\epsilon_1=10^{-3}$ we obtain a value of 
$F < 10^{-5}$ demonstrating that this setup is `safe', at least at tree-level, with respect to the electroweak constraints. Similarly, we note that as in the 4-D case for values of $\epsilon_1$ 
of this magnitude or smaller, the anomalous $g-2$ value of the muon\cite{Hagiwara:2017zod} will still remains unexplained.

\begin{figure}[htbp]
\vspace*{-1.5cm}
\hspace*{-0.1cm}
\centerline{\includegraphics[width=5.0in,angle=90]{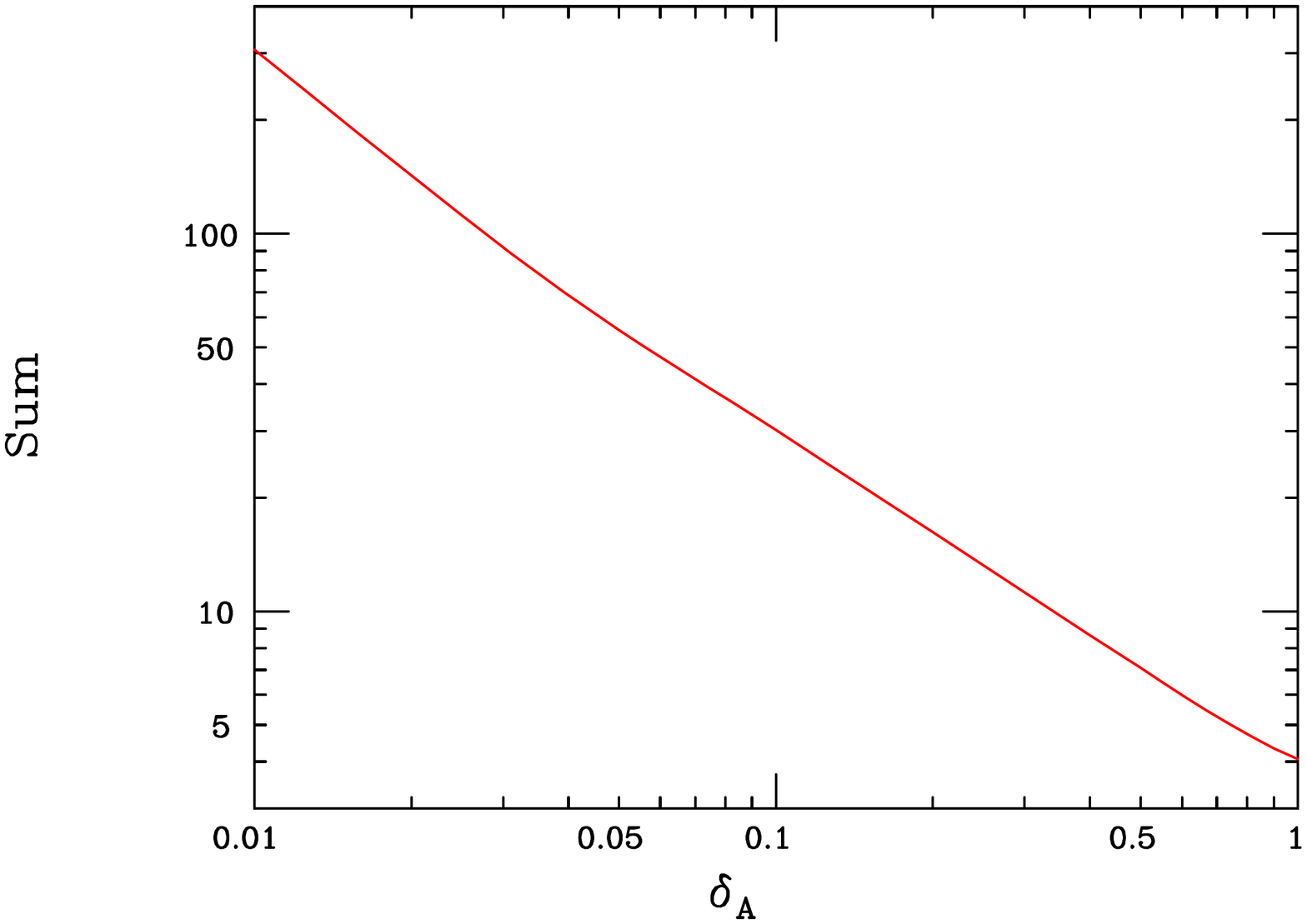}}
\vspace*{-1.50cm}
\caption{The dependence of the (normalized) sum over the $\epsilon_n^2$ as a function of the $\delta_A$ BLKT as described in the text.}
\label{sum}
\end{figure}

From the above discussion we know that $\delta_A \geq |\delta_0|$; can we also see that a minimum value is required from the KK decomposition. This we find by noting 
that the infinite sum, $\sum_n ~\epsilon_n^2/\epsilon_1^2$, encountered above, must be below some fixed  {\it a priori} value. Clearly, 
as $\delta_A$ decreases(increases) this sum will correspondingly increase(decrease) in magnitude and it will diverge completely as $\delta_A \to 0$. To address this 
question,  Fig.~\ref{sum} shows the value of this sum as a function of  $\delta_A \leq 1$. We observe in this Figure is that for any value of $\delta_A \gsim 0.25$ this sum is 
$\lsim O(10)$ or less. However,  as $\delta_A$ becomes very small the sum grows rapidly as an inverse power of $\delta_A$. We will restrict ourselves to the `natural' range of 
$\delta_A$ corresponding to values not too far from unity in the discussion that follows in which case the value of the sum is below $O(10)$ and thus well-behaved. We note that since 
$\delta_A \geq |\delta_0|$ and $\epsilon_1^2=\epsilon_5^2 (N^V_1)^2$ we can obtain the rather weak bound $\delta_A > \frac{\pi \epsilon_1^2}{2c_w^2}$.

\begin{figure}[htbp]
\centerline{\includegraphics[width=5.0in,angle=90]{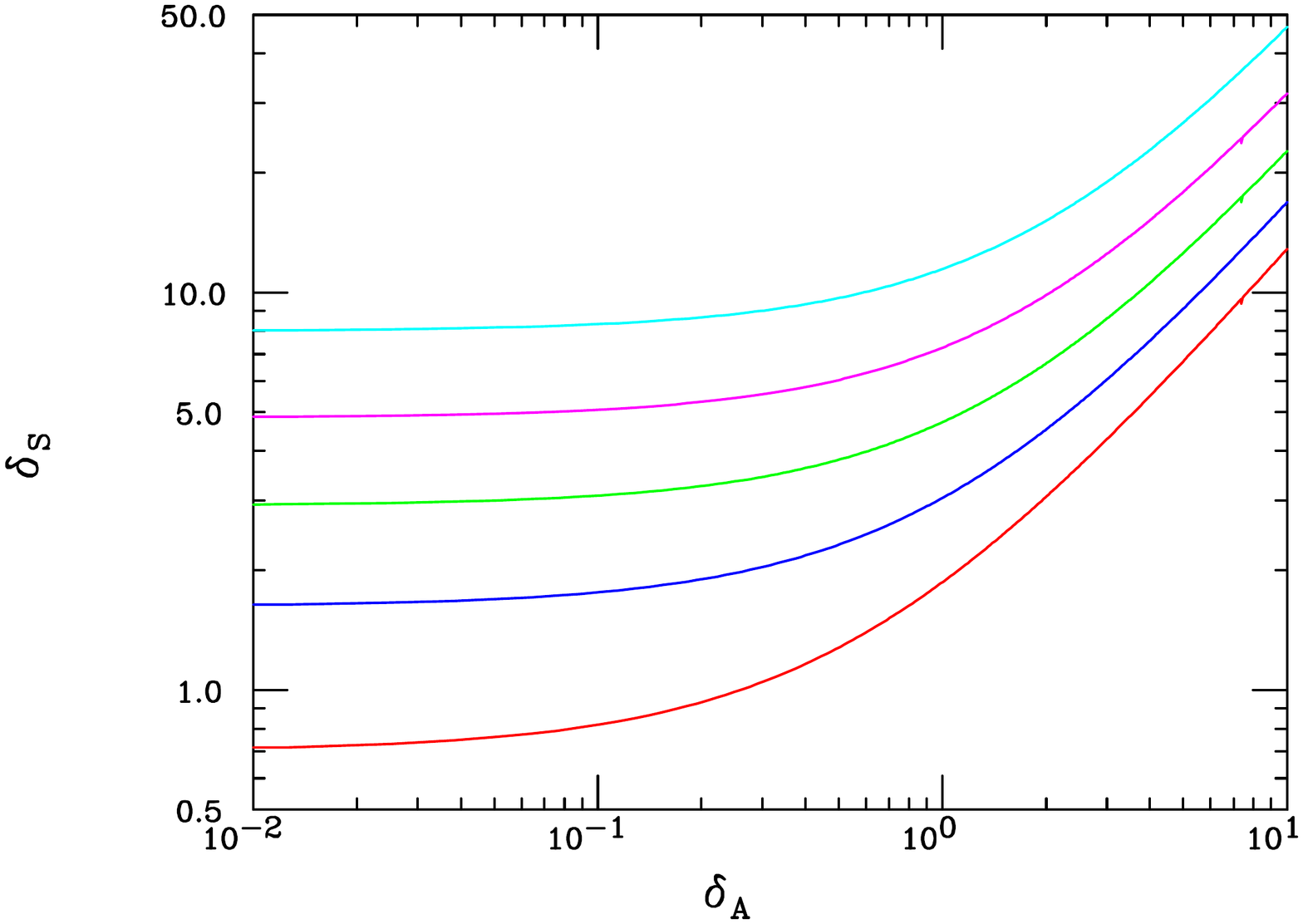}}
\vspace*{-3.1cm}
\centerline{\includegraphics[width=5.0in,angle=90]{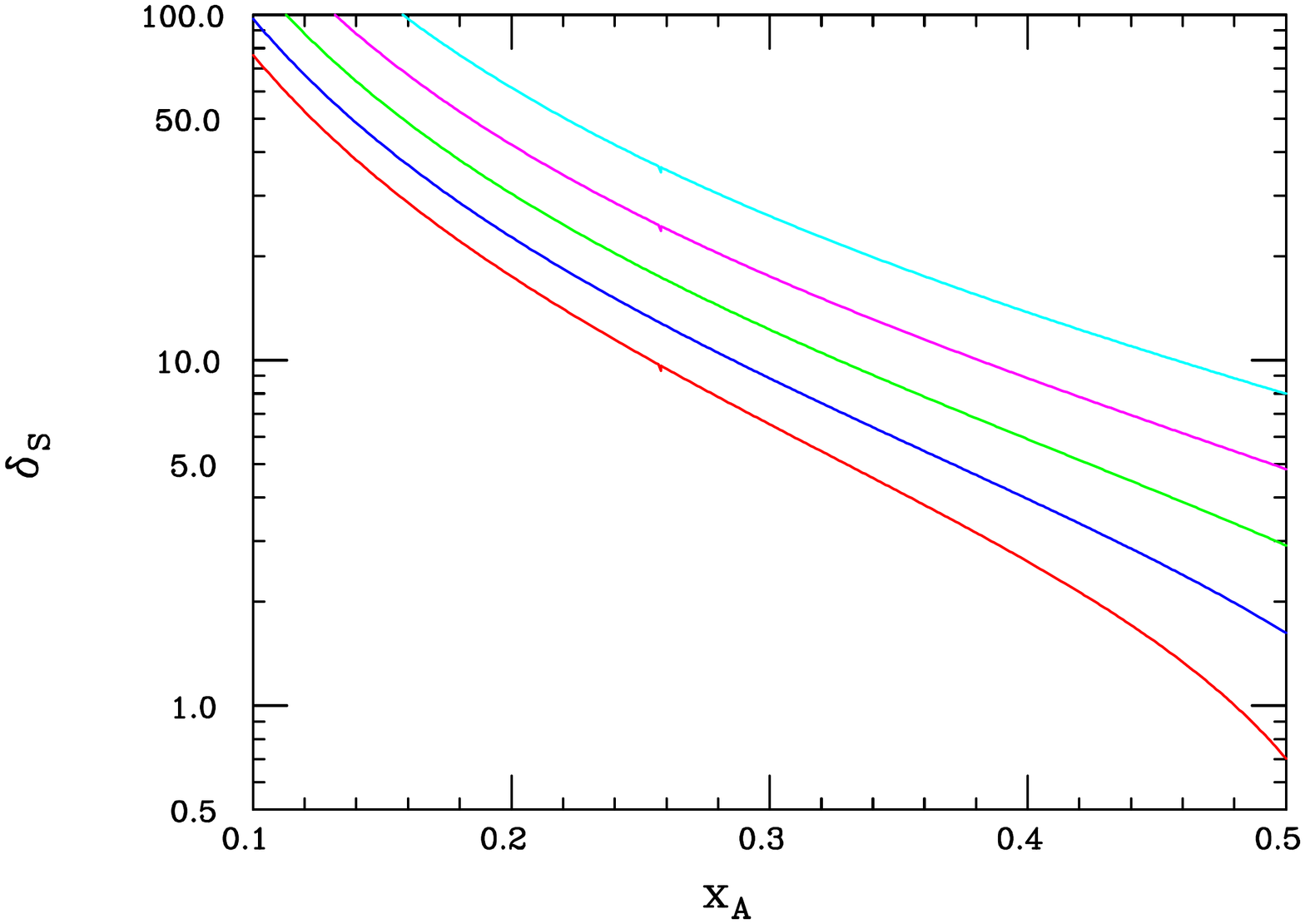}}
\vspace*{-1.90cm}
\caption{Top: Values of $\delta_S$ as functions of $\delta_A$ for (from top to bottom) $\lambda=0.5,~0.6,...,~0.9$. Clearly the $x^{V,S}_n$ values are identical when $\delta_A=\delta_S$. 
Bottom: Same as the Top panel but now as a function of the input value of $x_A=x^V_1$.}
\label{match}
\end{figure}

Although this gauge BLKT solves the KM consistency issue, the reduction in the the gauge KK masses, \ie, the values of the roots, seen in Fig.~\ref{masses}  now 
leads to a further issue since this BLKT, level by level, make these gauge fields lighter than the corresponding ones for the scalar DM fields. This 
we need to avoid in order to obtain a purely $p-$wave DM annihilation process. The natural solution  is to introduce a corresponding  BLKT, described by the 
corresponding dimensionless parameter $\delta_S$, for the scalar field $S$ on the $y=0$ brane (as the $s_n$ wavefunctions vanish at $y=\pi R$).  This new piece of the 
action is similar to the one for the gauge fields above:
\begin{equation}
S_{BLKT}'=\int ~d^4x \int_{0}^{\pi R} dy ~\Big[ (D_\mu S)^\dagger (D^\mu S)  \cdot\delta_SR~\delta(y) \Big]\,
\end{equation}
The equation governing the corresponding roots $x^S_n$ determining the scalar KK masses is exactly 
the same as that for the vector KKs above but with the trivial $\delta_A \to \delta_S$ replacement. In this case, we can write the $S$ wavefunctions as (noting that they still vanish 
on the $y=\pi R$ brane via the root equation)
\begin{equation}
s_n(y) =N^S_n \Big[\cos \frac{x^S_ny}{R} -\frac{\delta_S}{2}x^S_n \sin \frac{x^S_ny}{R}\Big]\,
\end{equation}
where $N^S_n$ is given by
\begin{equation}
(N^s_n)^2 = \frac{2}{\pi R}~\Bigg[1+\Big(\frac{\delta_S x^S_n}{2}\Big)^2+\frac{\delta_S}{2\pi}\Bigg]^{-1}\,.
\end{equation}
The behavior of the roots $x^S_n$ as functions of $\delta_S$ are identical to those of $x^V_n$ as functions of $\delta_A$ since they result from identical root equations. To 
obtain, \eg, $m_{S_1} < m_{V_1}$, one needs only require so that $\delta_S>\delta_A$.  Specifically, for a given value of $\delta_A$ (and thus for a specific $x^V_1$), to obtain a mass 
ratio  $m_{S_1}/m_{V_1}=\lambda <1$ we can obtain this result uniquely by choosing a specific value of $\delta_S$ given by 
\begin{equation}
\delta_S = \frac{2\cot \pi\lambda x^V_1}{\lambda x^V_1}\,
\end{equation}
as is shown in Fig.~\ref{match}. We observe that for small $\delta_A$ where the lowest gauge KK root hardly differs from 0.5, the required values of $\delta_S$ for any fixed value 
of $\lambda$ is essentially independent of $\delta_A$ and for values of $\delta_A \sim 1$ the required value of $\delta_S$ to obtain $\lambda=0.5$ grows significantly large , \ie, 
$\geq 10$ or so. Further, we see that if we allow $x^V_1$ to decrease significantly then, correspondingly, very large values of $\delta_S$ are required to obtain the typical 
$\lambda$ values of interest. 
This Figure also leads us to contemplate the region where $\lambda<0.5$ not directly considered in these plots and which requires substantially large values of $\delta_S$ implying 
that it might be disfavored. For these $\lambda$ values the decay mode $V_1\to S_1^\dagger S_1$ is open so that the lightest DP gauge KK state can decay to stable DM. Since 
we assume that $g_D \sim 0.1-1$ while $\epsilon_1 < 10^{-3}$, the branching fraction for this decay is {\it very} close to unity. A short consideration of this parameter  
space reveals that now all of the scalar DM and DP gauge KK states will eventually cascade down into a combination of the fields $S_1,S_1^\dagger$ and $V_1$. This implies that if 
$\lambda <0.5$ the decays of {\it all} of the tower fields will eventually lead to invisible final states.{\footnote {We note that once the values of $\delta_A$, $\lambda$ and $R^{-1}$, to set 
the overall mass scale, are chosen the entire phenomenology of this scenario is completely fixed.}}  This is quite different than the region $0.5 \leq \lambda \leq 1$ wherein these KK cascade 
decays can be rather complex, as will be discussed below. Since the production and decay of the gauge KK tower fields will end up producing 
missing energy in its various manifestations, it is interesting to ask whether the presence of such excitations would be observable {\it relative} to the corresponding signals anticipated in the 
4-D DP model with a similar mass hierarchy. While when $\lambda <0.5$ (but {\it not} very small) we might not expect any significant qualitative difference in the results 
for the thermal DM relic density or for the spin-independent scattering cross section, when the gauge KK states are directly produced at accelerators some 
differences with 4-D expectations might be observable. 

\begin{figure}[htbp]
\centerline{\includegraphics[width=5.0in,angle=90]{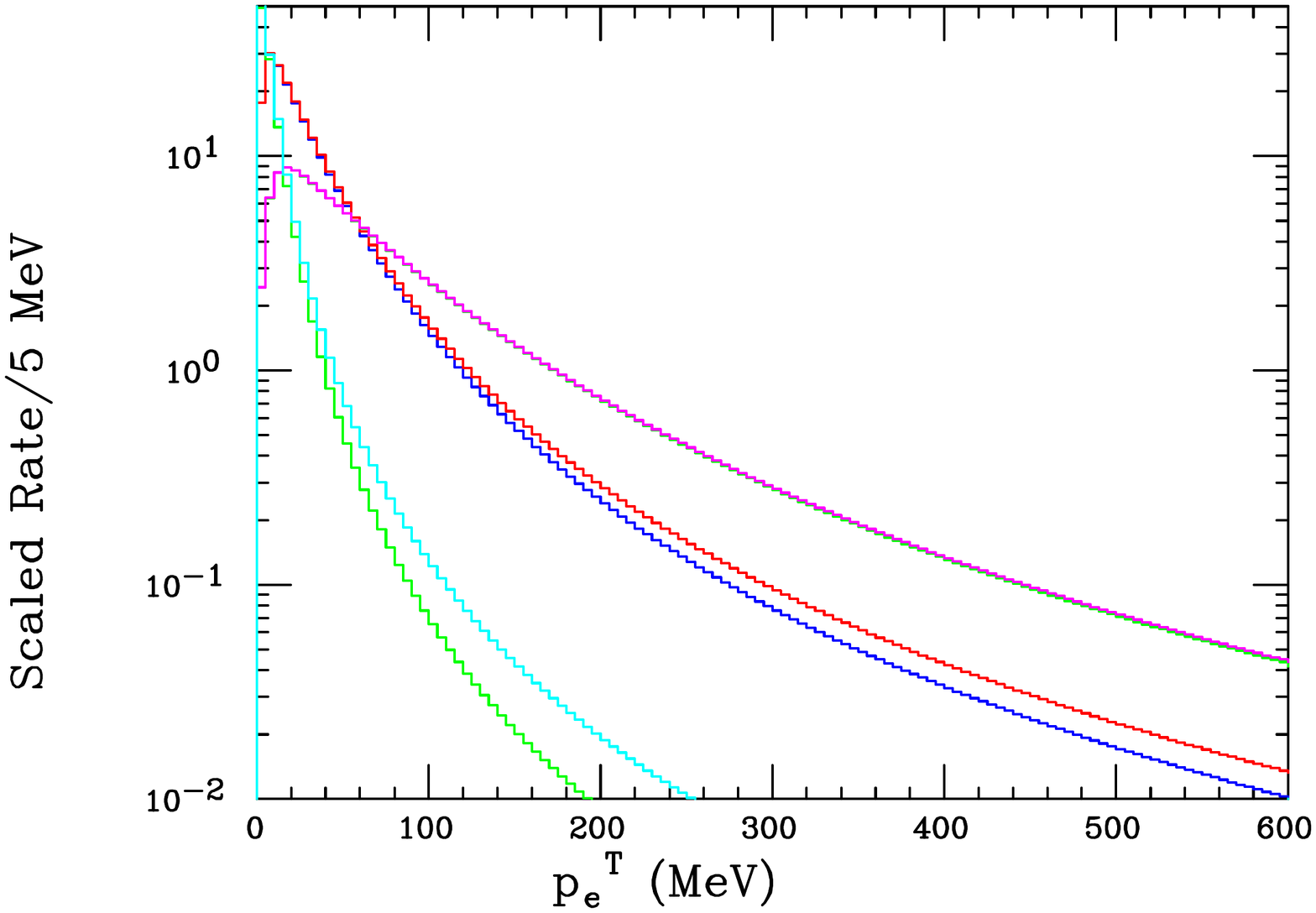}}
\vspace*{-3.1cm}
\centerline{\includegraphics[width=5.0in,angle=90]{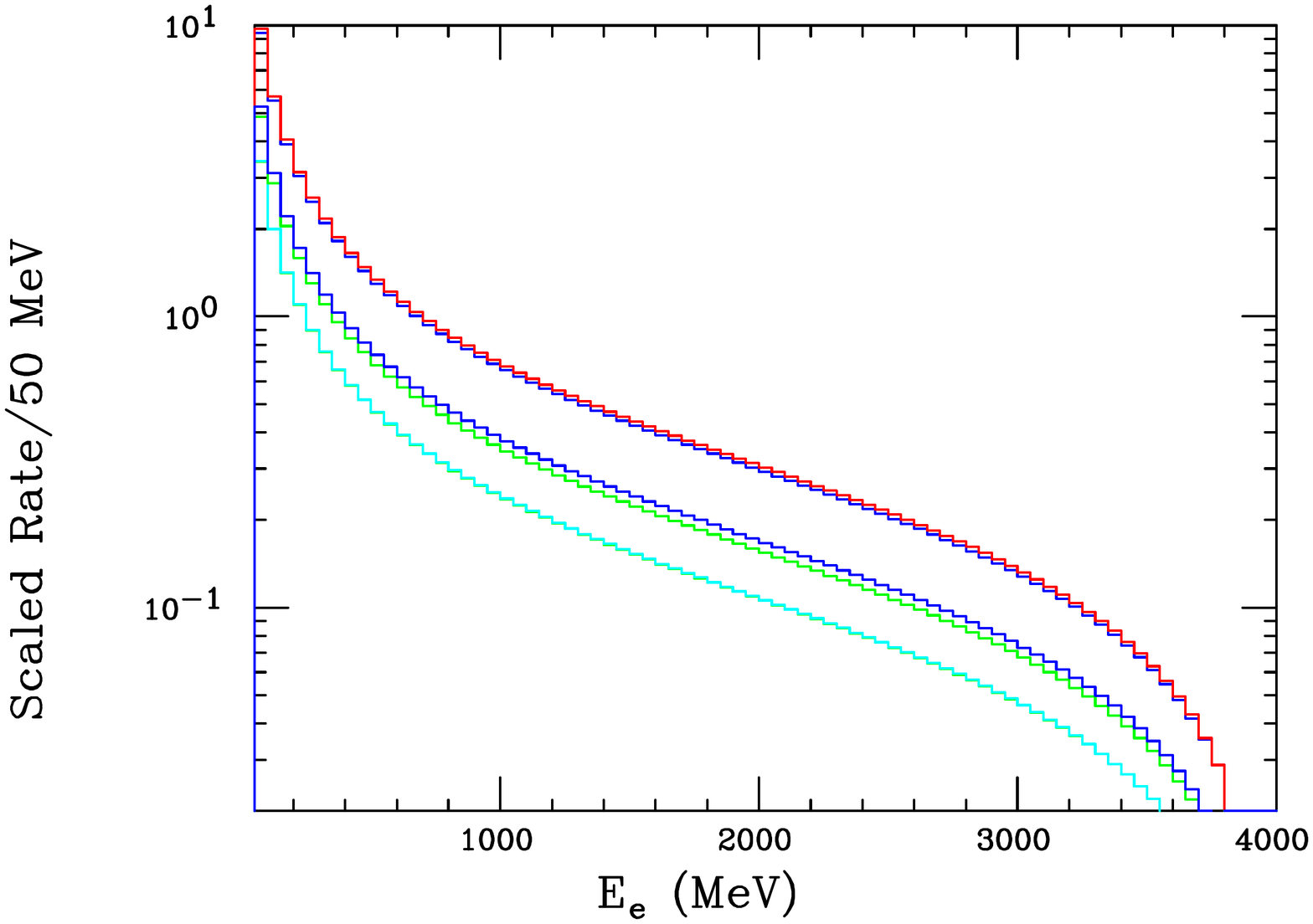}}
\vspace*{-1.90cm}
\caption{ Comparison 4-D and 5-D model predictions for the electron $p_T$ (top) and recoil energy (bottom) distributions at LDMX. In the upper panel in the center of the plot 
the histograms are, from top to bottom, for $R^{-1}=500$, 200 and 50 MeV, respectively.  In the lower panel, in the center of the plot the histograms are, from top to bottom, are
for $R^{-1}=200$, 50 and 500 MeV, respectively. Note that pair-to-pair normalizations are arbitrary.}
\label{LDMX}
\end{figure}

One of the best ways to probe this `all invisible' decay scenario is to generalize the 4-D analysis for the direct production of the invisibly decaying DP in, \eg, a fixed target 
experiment. To be specific, we consider the (approximately forward) production of a DP KK tower at the proposed LDMX experiment\cite{LDMX,Mans:2017vej} which will scatter an 
$E_0= 4-8$ GeV $e^-$ beam off of a Tungsten target. In such a case, once $\delta_A$ and $R$ are fixed, a tower of all of the kinematically accessible, on-shell, DP KK states is produced with a 
known rate all of which cascade down to invisible DM.  For example, the electron recoil energy spectrum in such a situation is obtained by complete analogy with the 4-D case and, in the 
Improved Weizsacker-Williams Approximation (IWW)\cite{Bjorken:2009mm}, is given by 
\begin{equation}
\frac{d\sigma}{dy} =8\alpha^3~\chi~\sum_n~\Bigg[\epsilon_n^2\beta_n~\frac{y+(1-y)^2/3}{(m^V_n)^2 y/(1-y) +m_e^2(1-y)^2}\Bigg]\,
\end{equation}
where here $\chi$ is a nuclear form factor\cite{Bjorken:2009mm}, $\beta_n^2=1-(m^V_n/E_0)^2$ and $y=E_e/E_0$ with $m_e(E_e)$ being the electron mass and recoil energy. Similarly  
one can generalize the expression for the electron $p_T$ distribution to the case of multiple contributing DP KK states. Since the effect of EDs will be the most significant when 
the KK couplings expressed through the $\epsilon_n$'s are large we will assume that $\delta_A=0.5$. Similarly, we expect that the ED effects will be largest the greater the number of KK states  
that can contribute to the above sum for a fixed value of $E_0$; this will happen when we chose $R^{-1}$ to be small. In Fig.~\ref{LDMX}, we pair-wise compare, for three different 
values of $R^{-1}$, the results of a 4-D IWW calculation for both the electron recoil spectrum and the scattered electron $p_T$ distribution with those obtained using the present 5-D model 
employing the cuts in the LDMX study\cite{LDMX}.  The 4-D model will be defined to be the same as the 5-D model when truncated to only the single lowest KK contribution to the cross section. 
Here, only pairs of curves for each $R^{-1}$ should be compared and the overall normalizations ignored since we are examining only the relative shape differences in these distributions as 
normalization shifts can be accounted for by a corresponding change in $\epsilon_1$. As expected, when $R^{-1}$ is 500 MeV, the 4-D and 5-D distributions for both observables are completely 
indistinguishable as only very few KK modes can contribute. As $R^{-1}$ decreases, to 200 MeV and then to 50 MeV, the KK contributions become more visible in the case of the $p_T$ distribution 
whereas the electron recoil energy spectrum appears to show little if any sensitivity to EDs.  Clearly, for even larger values of $R^{-1}$ there will be no ED sensitivity while, for fixed $R$, the 
sensitivity will grow with the value of the beam energy, $E_0$. Similarly, lowering $\delta_A$ will increase the KK tower couplings and result in an increase in sensitivity. It is beyond the scope of the 
present work to determine whether or not experiments such as LDMX could be sensitive to these differences as this will require a much more sophisticated study including detector simulation.

As is well-known\cite{Battaglieri:2017aum}, significant additional sensitivity to gauge KK production with invisible decays is possible from radiative production in the decay of mesons, 
\eg, $\pi,\eta \to \gamma V_i$ and in $e^+e^-$ collisions at, \eg. BELLE-II\cite{Belle}, although in both cases multiple gauge KK states may now become accessible via the photon recoil 
spectrum. Within our framework, the lightest gauge KK state may or may not decay to SM fields while the next set of KK's will decay solely to DM and its KK excitations as will be 
discussed below. This production of multiple photon peaks in the decay or recoil spectrum would provide a rather unique signal for the present scenario.

The direction production of an extended DP sector may also be observable using other techniques such as in the mixing-induced SM Higgs decays $H\to ZV_i,V_iV_j$ as 
discussed above. Once we chose the values of $\delta_A$ and $R^{-1}$ these decay rates can be calculated employing the toolbox we have developed. However, 
we immediately find that the current rough constraint $B(H\to ZV_i,V_iV_j)<0.25$, corresponding to a partial width $< 1$ MeV, is very easily satisfied for the suggestive choices 
$\delta_A=0.5$, $R^{-1}=100$ MeV and $\epsilon_1=10^{-3}$. In the case of $H\to ZV_i$, this parameter choice yields a partial width 
of $\simeq 0.16$ eV while for the double mixing suppressed mode $H\to V_iV_j$, we obtain a partial width of only 
$\simeq 1.5\cdot 10^{-5}$ eV. Note that increasing the value of $\delta_A$ makes the $\epsilon_n$'s fall even faster so the KK sums will decrease while changing $R^{-1}$ sets 
the mass scale of KK modes contributing to these sums; $\epsilon_1$ remains as an overall scaling factor. We can safely conclude that these Higgs decay modes are far beyond the 
reach of experimental detection for reasonable parameter choices.

Once the wavefunctions and masses of the vector and scalar KK states are determined one can calculate all of the couplings of interest, in particular, the explicit values of 
the couplings of the KK scalars to the KK gauge tower states as given by the quantities
\begin{equation} 
c_{mn}^i = \int_0^{\pi R} ~s_n(y)s_m(y)v_i(y)~dy\,
\end{equation}
as described earlier.  As noted the actual physical couplings, in terms of 4-D coupling $g_D=g_{5D}c_{11}^1$ (with $\alpha_D=g_D^2/4\pi$),  are given by the ratio 
$\tilde c_{nm}^i=c_{nm}^i/c_{11}^1$.  With these in-hand we now turn to some DM phenomenology. Of particular interest is how the lightest scalar (and 
its complex conjugate) which forms the DM interacts with the DP KK tower, \ie, the couplings $\tilde c_{11}^i$. Knowing these quantities 
and the KK masses we can calculate, \eg, the DM annihilation rate which determines the relic density as well as the DM spin-independent direct detection (SI-DD) cross section 
up to a common overall factor related to the specific KK mass scale. In both cases, when we calculate the amplitude for the relevant process, we must perform a KK sum over an 
infinite tower of the gauge KK states; we first turn to the SI-DD process.{\footnote {We note that in a general gauge the unphysical $V_5$ field does not mediate any interaction between the 
scalar sector and the SM fields in the present setup.}}

As in 4-D, the SI scattering process occurs via $t$-channel exchange of dark photons, but now with the full KK tower of states participating. For the light DM that we consider 
here, $\sim 10-100$ MeV, scattering off of electrons is likely to provide the greatest sensitivity\cite{Battaglieri:2017aum}. Assuming a form factor of unity since $m_{DM}^2>> m_e^2$ , 
this cross section can be expressed as (for either $S$ or $S^\dagger$ scattering)\cite{Chen:2016tdz,Essig:2017kqs} 
\begin{equation} 
\sigma_e = \frac{4\alpha \mu^2 g_D^2 \epsilon_1^2}{(m^V_1)^4} ~\Bigg[\sum_n ~(-1)^{n+1} \frac{\epsilon_n}{\epsilon_1}~\tilde c_{11}^n ~\frac{(m^V_1)^2}{(m^V_n)^2}\Bigg]^2\,
\end{equation}
where $\mu=m_em_{DM}/(m_e+m_{DM})$ is the reduced mass $\sim m_e$  for the DM masses of interest to us; here, $m_{DM}=m^S_1$. Numerically this yields the result 
\begin{equation} 
\sigma_e \simeq 3.0\cdot 10^{-40} \rm{cm}^2 ~\Big(\frac{100 \rm{MeV}}{m^V_1}\Big)^4 ~\Big(\frac{g_D\epsilon_1}{10^{-4}}\Big)^2 ~\times \rm{Sum}\,
\end{equation}
whereby the quantity `Sum' represents the squared KK summation of the previous expression which we expect to be $\sim O(1)$  as the series converges rapidly and whose numerical value 
we will return to shortly. Note that here `Sum' isolates the {\it only} difference between the prediction of this ED scenario and the more familiar one in the 4-D case. For representative parameter 
values SuperCDMS is likely to be able to probe this range of cross sections in the future but it now lies a few orders of magnitude below the current constraints\cite{Essig:2015cda}.

The calculation of the thermal DM annihilation cross section into final state electrons (the most likely possible final state for typical DM masses) can be expressed in a similar 
fashion by writing $\sigma v_{rel}=\tilde bv_{rel}^2$, where the detailed kinematic information, including the sub-leading terms in the velocities, and (away from any resonances for simplicity) 
is contained in the parameter $\tilde b$ which is given in the limit of a zero electron mass by\cite{Chen:2016tdz,Berlin:2014tja}
\begin{equation} 
\tilde b=\frac{g_D^2 e^2 \epsilon_1^2}{192\pi m_{DM}^2} ~\frac{\gamma^4}{\gamma^2-1}~ \sum_{n,m} (-1)^{n+m} \Bigg[\frac{(\epsilon_n\epsilon_m/\epsilon_1^2)~\tilde c_{11}^n\tilde c_{11}^m}
{(\gamma^2-r_n)(\gamma^2-r_m)}\Bigg] \,
\end{equation}
where here the double sum is over the gauge KK tower states, $\gamma^2=s/4m_{DM}^2$ is the usual kinematic factor determined by the DM velocities employing the standard 
Mandelstam variable and  $r_n=(m^V_n)^2/4m_{DM}^2$. To be specific below we will assume the freeze-out temperature to be $x_F=m_{DM}/T\simeq 20$ so that at freeze-out 
$<v_{rel}^2>\simeq 0.3$. In the case where other final states, such as muons and light hadrons, can also contribute, this cross section must be corrected by the well-known 
$R$-ratio factor. As we'll see from the mass spectra of our KK states in Table~\ref{decay1} below, for our benchmark models we not only have  $2m^S_1 > m^V_1$ but also that 
$2m^S_1$ is significantly below $m^V_2$ implying that the thermal DM annihilation cross section is dominated by phase space regions far from any of the narrow $s$-channel 
KK resonances.

\begin{figure}[htbp]
\centerline{\includegraphics[width=5.0in,angle=90]{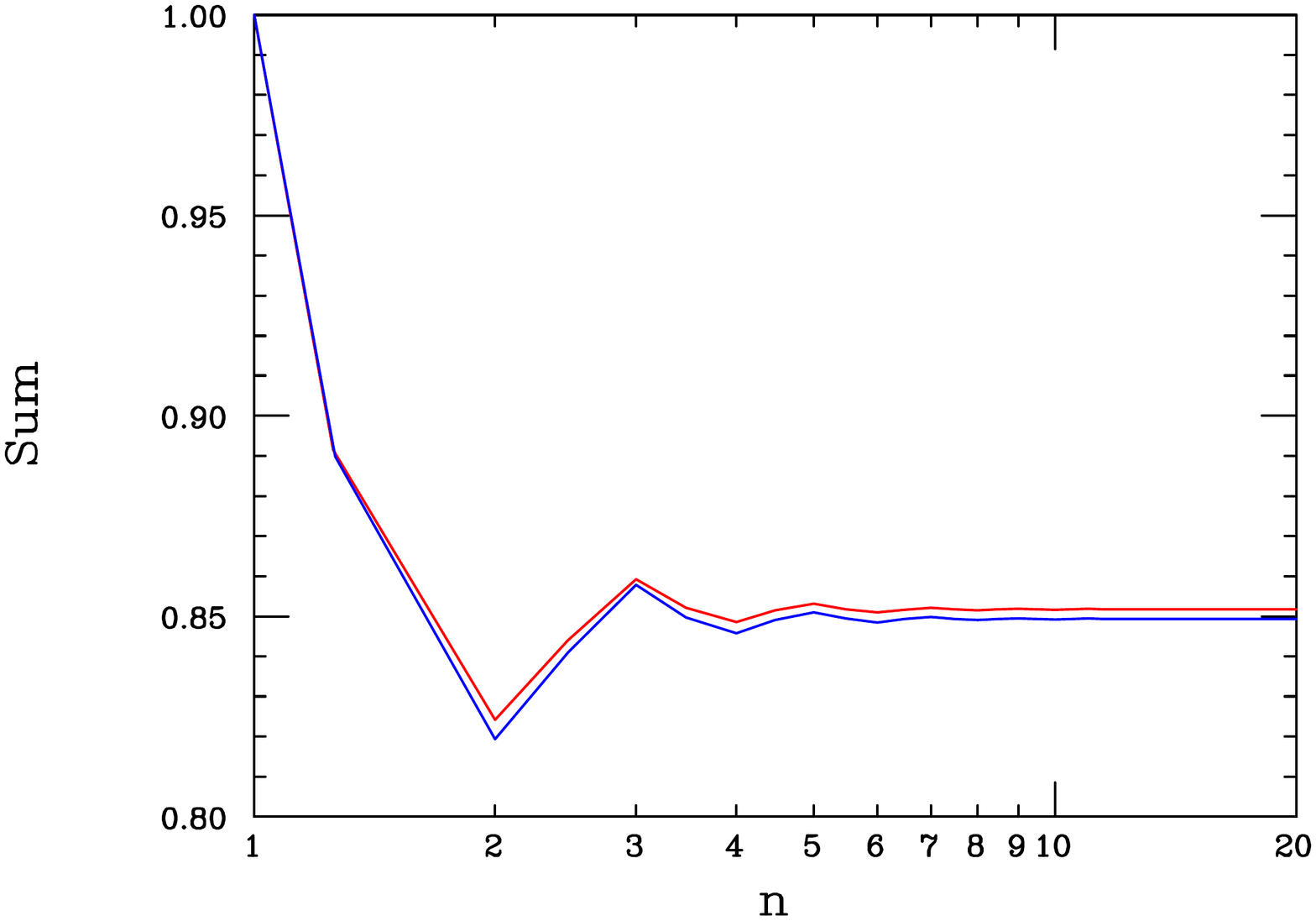}}
\vspace*{-3.1cm}
\centerline{\includegraphics[width=5.0in,angle=90]{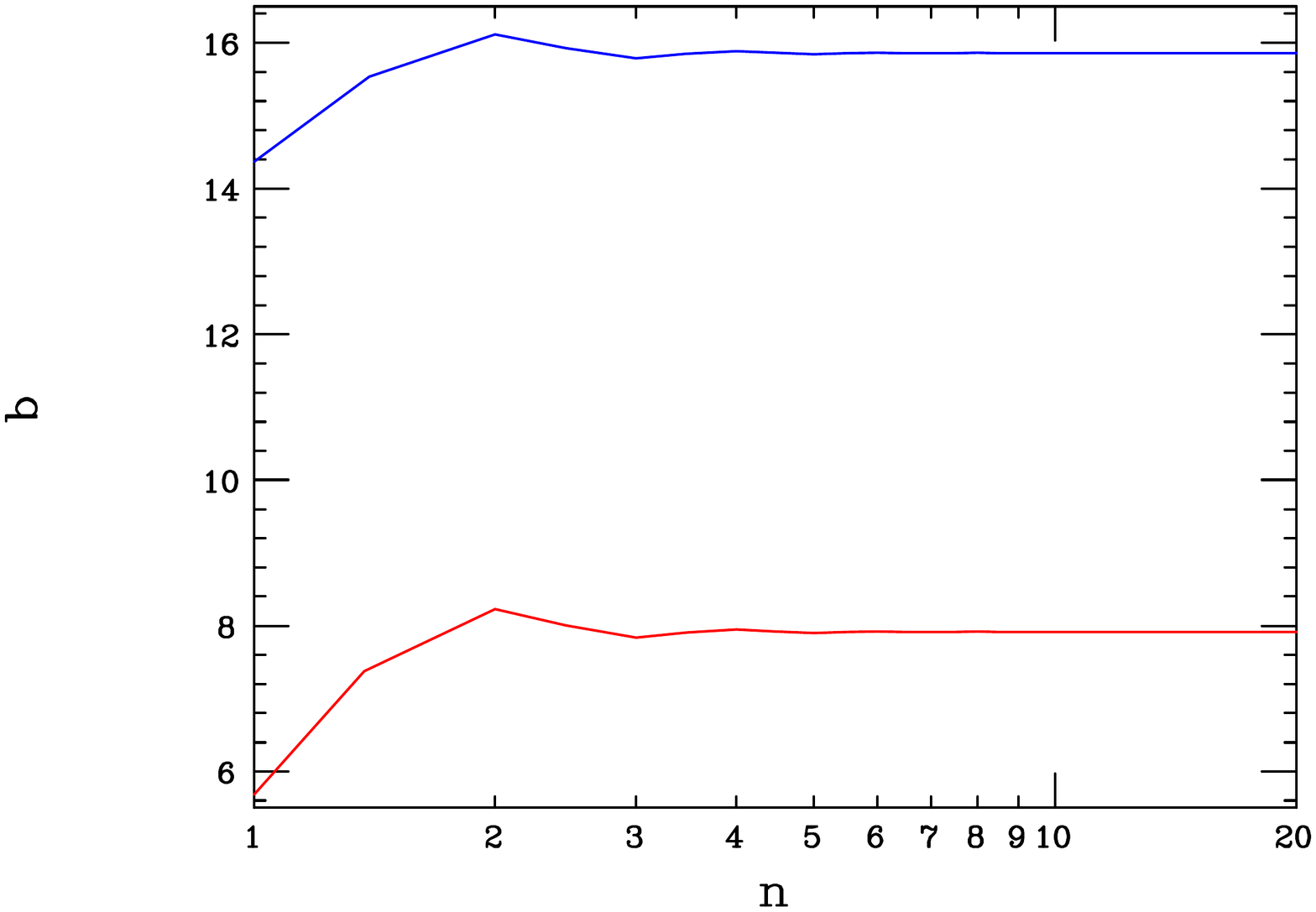}}
\vspace*{-1.90cm}
\caption{Top: Values of the quantity `Sum' appearing in the expression for the thermal DM annihilation cross section as a function of the number of gauge KK tower states, $n$, 
included in the sum,  as described in the text; the upper(lower) curve corresponds to the case of BM1(BM2).  Bottom:  value of the quantity $b$, as defined in the text, for BM1 
(upper curve) and for BM2 (lower curve).}
\label{dm1}
\end{figure}

To go further in our numerical evaluations of both the SI cross section and $\sigma v_{rel}$, we need to choose some specific benchmark models (BM) forcing us into some  
particular parameter choices. Here we give two examples both of which have $\delta_A=0.5$. For BM1, we take $m^V_1/m_{DM}=0.8$ implying $\delta_S\simeq 2.38$, while 
for BM2, we assume that $m^V_1/m_{DM}=0.6$ implying $\delta_S\simeq 6.03$. Apart from the overall mass scale set by $R^{-1}$, these quantities determine the 
complete model phenomenology.  In the upper panel of Fig.~\ref{dm1} we find the value of the quantity `Sum' defined above for our two benchmark points as a function of the 
number of contributing gauge KK tower states $n$. Here we see that ($i$) the results for these two BM points are essentially identical, ($ii$) the KK summation converges very 
rapidly, roughly by the time the $n\sim 5$ KK state is reached, to its final value. ($iii$) The value of this sum is less than unity due to the destructive interference among the gauge KK 
exchanges, \ie, `Sum' $\simeq 0.852(0.849)$ for BM1(BM2). This means that the entire KK tower above the lowest level only makes a $\sim 7\%$ contribution to the amplitude. 
Finally, ($iv$) we see that the 4-D and 5-D predictions are numerically quite close. It is important to emphasize 
the very rapid convergence of these sums and the essentially negligible contributions of the higher KK states here.

The lower panel of Fig.~\ref{dm1} shows the values of a quantity $b$ for both BM points; in this Figure we have rescaled the quantity $\tilde b$ above by an 
overall factor so that this quantity $b$  as shown here is both dimensionless and is roughly $O(10)$:  
\begin{equation}
\tilde b =b ~\Bigg[\frac{g_D^2e^2\epsilon_1^2}{m_{DM}^2 \rm {(GeV^2})}\Bigg] ~10^{-20} \rm{cm^3 s^{-1}}\,.
\end{equation}
In this panel we see that BM1(BM2)  leads to a value of $b \simeq 7.9(15.9)$ which differ by roughly a factor of $\sim 2$ due to the mass spectrum and various coupling 
variations. Note that, roughly, for $g_D\epsilon_1 \sim 10^{-4}$ and $m_{DM} \sim 10-100$ MeV we can easily obtain a thermal cross section of $\sim 9\cdot 10^{-26}~\rm{cm^3s^{-1}}$ 
as needed to reproduce the observed relic density for light complex DM masses\cite{Steigman:2015hda}{\footnote{We remind the reader that for DM that is not self-conjugate the 
required annihilation cross section is twice that of the the canonical value.}}  It is again important to emphasize 
the very rapid convergence of these sums and the essentially negligible contributions of the higher KK states beyond $n\sim 5$ here.

Next, we consider the decay properties of the various $S_n$ and $V_n$ states. By construction, for both BM points, $S_1$ and $S_1^\dagger$ are stable states forming 
the DM while $V_1$ decays only into SM final states as the decay $V_1\to S_1^\dagger S_1$ is kinematically forbidden; furthermore, $V_2$ only decays into 
$S_1^\dagger S_1$ since $g_D^2>>(e\epsilon_1)^2$.  In this 5-D model, $V_1$ acts like the 4-D DP decaying to only SM states while $V_2$ acts like the 4-D model 
where the DP decays only to DM. In a similar fashion, the decay $S_2\to S_1V_1$ occurs with a 100\% branching fraction.  The decays of 
the higher KK states are found to be somewhat sensitive to the BM choice due to the differences in couplings and phase space although the gauge KK masses 
are the same for both BMs. To see these mass differences explicitly, Table~\ref{spec} summarizes the masses of the lowest lying KK states for both BMs in units of $R^{-1}$. Here 
we see that as we ascend up the scalar KK tower the mass differences between the two chosen BM points vanishes as might have been expected based. In addition, we observe 
that the gauge and scalar KK states also eventually will become degenerate for large $n$, as expected. 
\begin{table}
\centering
\begin{tabular}{|l|c|c|c|} \hline\hline
KK level&     V      &  S(BM1)    &   S(BM2)   \\
\hline
~~~~1   &   0.463     &   0.371    &  0.278    \\
~~~~2   &   1.393     &   1.198    &  1.094     \\
~~~~3   &   2.332     &   2.123   &   2.051    \\
~~~~4   &   3.281     &   3.087   &   3.035    \\
\hline\hline
\end{tabular}
\caption{Masses of the four lightest gauge and scalar KK states in units of $R^{-1}$ for the two BM points considered in the text.}
\label{spec}
\end{table}
Finally,  we examine the branching fractions for the various gauge mediated decay modes of the heavier KK states for the two BMs; these are found in Table~\ref{decay1}. 
The decay width for $S_n(m)\to S_m(m')V_l(m_V)$ (with the relevant masses in the parentheses) is given by
\begin{equation}
\Gamma_S=\frac{g_D^2 (\tilde c_{nm}^l)^2 m^5}{16\pi m_V^4} ~\Bigg[1-2\frac{m_V^2+m'2}{m^2}+\frac{(m_V^2-m'^2)^2}{m^4}\Bigg]^{3/2}\,.
\end{equation}
Correspondingly, for the decay $V_i(m_V)\to S^\dagger_j(m_j)S_k(m_k)+$h.c., we find a similar expression
\begin{equation}
\Gamma_V=\frac{g_D^2 (\tilde c_{jk}^i)^2 m_V}{24(1+\delta_{jk})\pi} ~\Bigg[1-2\frac{m_j^2+m_k^2}{m_V^2}+\frac{(m_j^2-m_k^2)^2}{m_V^4}\Bigg]^{3/2}\,.
\end{equation}

In the Table we see that there can be quite significant differences in how the various KK states decay based on the small differences in masses and the variations in the $\tilde c_{nm}^i$ 
couplings. Clearly, searches for these more massive KK states will be influenced by these parametric variations. The fact that these two BMs can show such differences suggests 
that even greater variations are likely possible as we scan over the full parameter space.  As noted, once decays of these light KKs into other dark sector states are kinematically 
allowed the corresponding lifetimes are generally controlled by the coupling factors $\sim g_D^2 \times ~O(1)$  so that such decays are quite rapid. Of course the lightest KK gauge 
state, which decays to SM fields via $(e\epsilon_1)^2$ can be long-lived as has been often discussed in the literature\cite{Davoli:2017swj} for the 4-D case with typical $c\tau$ 
values of order 100 $\mu$m for $\epsilon_1 \sim 10^{-4}$ and masses of $\sim 100$ MeV. As we progress up the various KK towers, decay widths will increase due to the usual phase 
space and mass factors although in most cases these will be compensated for by shrinking values of the relevant parameters $\tilde c_{nm}^i$ and possible phase space suppressions. 

In any given experiment where multiple KK states $V_n$'s are produced on-shell by interactions, cascade decays through all of the lower mass KK states will occur with 
various BFs which we've seen are parameter dependent. However, in all cases where $m^S_1/m^V_1$ is in the $0.5-1$ range considered here, this cascade will produce a rather 
complex final state: multiple DM scalar pairs, appearing as as missing energy, as well as potentially numerous $V_1$'s,  which all produce relatively soft-lepton pairs, possibly 
with displaced vertices, depending upon the precise values of $R^{-1}$ and $\epsilon_1$.  This complex cascade phenomena is a rather unique signature of the present scenario 
and warrants further detailed study.

\begin{table}
\centering
\begin{tabular}{|l|c|c|c|} \hline\hline
~~~Process             &    BF(BM1)    &   BF(BM2)   \\
\hline
$S_3\to V_2S_1$  &  1.20     &  0.62 \\
$S_3\to V_1S_1$  &   5.10    &  1.78   \\
$S_3\to V_1S_2$  &   93.7     &  97.6   \\
    &     &     \\
$V_3\to S_1^\dagger S_1$  &  74.9     &   97.3    \\
$V_3\to S_1^\dagger S_2$+h.c.  &   25.1     &  2.71   \\
   &     &  \\
$V_4\to S_1^\dagger S_1$ &  45.9     & 39.5      \\
$V_4\to S_1^\dagger S_2$+h.c. &   51.5    & 18.9       \\
$V_4\to S_2^\dagger S_2$  &  1.67      & 38.8    \\
$V_4\to S_3^\dagger S_1$+h.c.  &  0.95     &  2.81     \\
\hline\hline
\end{tabular}
\caption{Branching fractions for the various decay modes in per cent for the next highest gauge and scalar KK states in both BM scenarios discussed in the text.}
\label{decay1}
\end{table}

As a final note, we now calculate the partial width of the $Z$ into dark sector fields\cite{nikita},  \ie, the tower of scalars, 
$Z\to \sum_{n,m} S^\dagger_nS_m+$h.c.,  which is induced by the mixing of the $Z$ with the $V_i$ as previously described.  As discussed above this may provide an important 
additional constraint on the model parameter space particularly if all the subsequent decays end up as missing energy. Here one must also assume a value for the 
(inverse) compactification radius, $R^{-1}$ as mentioned earlier since this sets the overall mass scale of the KK states. The smaller $R^{-1}$ is chosen to be the greater the 
number of KK states that are kinematically accessible in this decay process. As noted, in much of the parameter space these modes will appear as additional contributions to the 
partial width for $Z\to$ invisible (since the cascade products are quite soft or will explicitly be missing energy) which is bounded to be below $\sim 1-2$ MeV[\cite{pdg}. To be definitive, 
we choose  representative values of $R^{-1}=100$ MeV, $g_D \epsilon_1=10^{-4}$ and the couplings and mass spectra corresponding to BM2 for our numerical example. Here we 
sum over the set of kinematically accessible final states although those higher up in the tower make only an infinitesimal contributions due to their highly suppressed couplings. 
With these input values we find a partial width of $\sim 0.02$ MeV which is more than a factor of $\sim 50-100$ below the current experimental 
constraint and so presents no phenomenological issues.  Here we see that the suppression from the shrinking values of the $\epsilon_n$ and $\tilde c^i_{jk}$ as the KK towers are ascended
allows predictions to be easily compatible with experiment. 

\begin{figure}[htbp]
\vspace*{-1.5cm}
\hspace*{-0.1cm}
\centerline{\includegraphics[width=5.0in,angle=90]{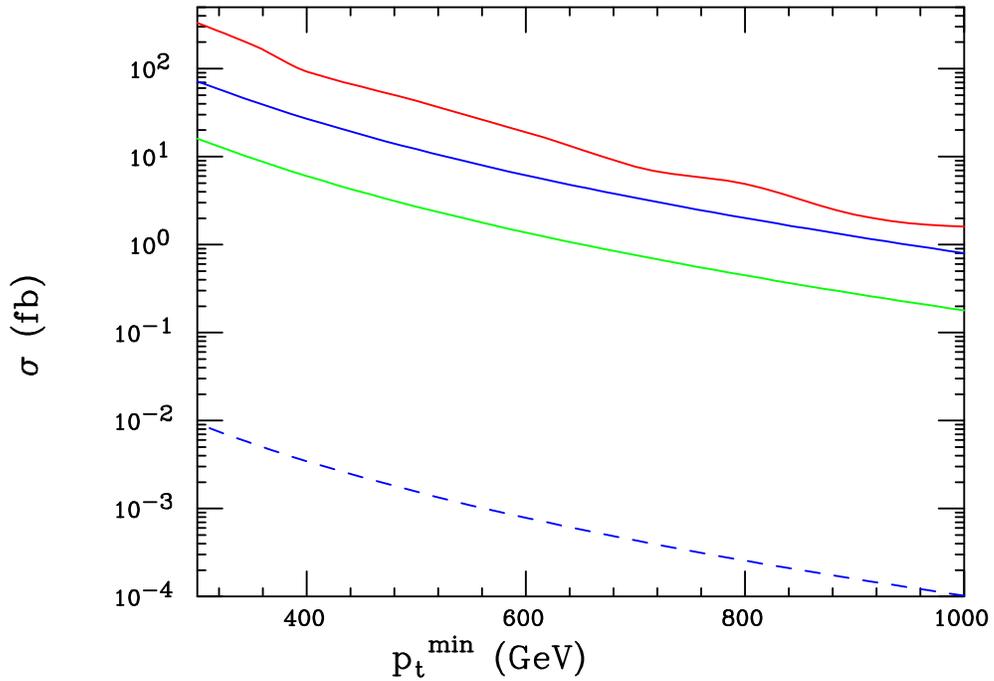}}
\vspace*{-1.50cm}
\caption{Comparison of the 13 TeV ATLAS `monojet' cross section upper limit as a function of the leading jet minimum $p_T$  (top red curve) with the predictions of several 5-D 
scenarios assuming that $\epsilon_1=10^{-4}$. The solid blue (green) curve is the prediction assuming $R^{-1}=100(200)$ MeV without any BLKTs. The lower blue dashed curve 
shows the resulting shift in the case of $R^{-1}=100$ MeV case when $\delta_A=0.5$ is assumed.}
\label{compare}
\end{figure}

Another example of this is very strong suppression of couplings due to the BLKTs is provided by the incoherent production of multiple gauge KK states at the the LHC in the `monojet'  
channel\cite{Liu:2017jzn} assuming that they all eventually decay to DM or soft decay products as is described above. If $R^{-1}=100(200)$ MeV then at $\sqrt s=$13 TeV many 
KK states are kinematically accessible although their couplings rapidly grow infinitesimal making for a vanishingly small contribution to the signal.
We follow the ATLAS monojet analysis\cite{mono} here and employ the mass dependence of the gauge KK couplings to quarks as obtained above, \ie, $\sim Q$ for low masses and 
$\sim Y$ for larger masses although this does not have a 
qualitative impact on our numerical results. Without any BLKT-induced suppression the expected rates for this process could be quite large. Fig.~\ref{compare} shows a 
comparison of the current ATLAS 13 TeV limit limit on this `monojet' signal as a function of the minimum $p_T$ of the leading jet. Also shown are the predictions of the present model in 
the absence of the BLKTs assuming $\epsilon_1=10^{-4}$ and with $R^{-1}=100(200)$ MeV. Even with this small value of $\epsilon_1$ we see that the predictions would not lie very 
far from the current ATLAS upper limit. However, assumimg that $\delta_A=0.5$, then the predicted signal rate falls by 
a factor of $10^{3-4}$, \ie, it lies very far from the current limit. Increasing our chosen value of $\delta_A$ will result in a greater suppression of the rate 
for this process since the $\epsilon_n$'s will fall off more quickly with increasing $n$.

\subsection{Model 2: Complex Scalar DM With a VEV}

When the SM bulk singlet field $S$ acquires a non-zero vev, $v_s$, the phenomenology of the model becomes more complex as there are more distinct physical states in 
the spectrum. In 4-D, the scalar vev is a necessary ingredient in order to break the $U(1)_D$ gauge symmetry and to give the DP a mass. As we saw in the preceding subsection, 
this vev was not needed for this purpose as, in the unitary gauge, the vector fields eat their 5th components, $V_5$, to acquire masses with these 
BCs. {\it However}, if it is non-zero, this vev will contribute to the DP and scalar KK tower masses. When this vev is non-zero, the complex field $S$ also decomposes 
(assuming CP conservation) into a set of CP-even ($h$) and CP-odd ($\chi$) fields; this CP-odd component is eaten by the Higgs mechanism in 4-D. Here 
one linear combination of the $\chi$ and $V_5$ towers becomes the KK Goldstone bosons generating the KK gauge masses whereas the other 
linear combination is realized as a physical CP-odd scalar KK tower. The magnitude of this admixture of these weak eigenstates is level-dependent and is partially 
determined by the dimensionless product $g_Dv_sR$. This scenario is very similar to that of the familiar 5-D Abelian Higgs model, described in detail in\cite{5d} and whose analysis we 
will follow and whose results we directly quote. However,  the present scenario differs in detail from this presentation in several ways: ($i$)  the the location of the SM fields is at 
$y=\pi R$, ($ii$) the ED interval is restricted to $0\leq y \leq \pi R$, ($iii$) our choice of BCs will produce non-zero masses even for the lightest KK 
modes, and, finally, ($iv$) the presence of a gauge BLKT on the SM brane is introduced as in the previous subsection to render the `undoing' of the KM on the SM brane physical.

The simplest version of this setup as is described below has a significant flaw in that the DM field, the lightest CP-even scalar,  can have too short of a lifetime to be the true 
DM; we will investigate how this problem can be evaded without any significant phenomenological changes in the next subsection.

The first step in the present analysis is to examine the impact once $S$ obtains a non-zero vev, \ie, once $S\to (v_s+h+i\chi)/\sqrt 2$ at the 5-D level, due to minimizing the potential:  
\begin{equation}
V_S= -\mu_S^2 S^\dagger S +\lambda_S (S^\dagger S)^2\,,
\end{equation}
which we re-express it in terms of the (almost) physical fields $h,\chi$ that will be KK expanded after application of the appropriate BCs. 
At quadratic order, the $h$ field obtains a 5-D bulk mass, $M_h^2=2\lambda_S v_s^2$, while the $\chi$ field (the would-be Goldstone in 4-D) remains massless in 5-D. The gauge 
field also obtain a bulk mass term, $M_V^2=(g_D Q_D v_s)^2$ where $Q_D$ is the dark charge of $S$ and is taken to be unity without loss of generality. 
Note that with our choice of BCs,  the bulk gauge mass is {\it not} the mass of the lightest gauge KK mode.  We assume that $h$ and $\chi$, satisfy the BCs 
$\partial_y (h_n,\chi_n)=0 $ at $y=0$ and that $h_n,\chi_n=0$ at $y=\pi R$, similar to that of $S$ in the previous subsection, and for simplicity without any associated 
BLKT (but which we will return to below); thus $h_n,\chi_n \sim \cos \sigma_n y$. Assuming CP conservation, the $h_n$ are realized as physical states with masses given by 
$(m^h_n)^2=(\frac{n+1/2}{R})^2+2\lambda_Sv_s^2$ and with the wavefunctions $h_n(y)= \sqrt{\frac{2}{\pi R}} \cos \frac{(n+1/2) y}{R}$. 
As mentioned earlier, the $V_{5n}$, which also experience the bulk gauge mass, and $\chi_n$ fields mix to form the KK Goldstone bosons, $G_n$, and the physical CP-odd scalars, 
$a_n$ as given by\cite{5d}  
\begin{eqnarray}
G_n &=& \frac{\sigma_n V_{5n}+g_Dv_s \chi_n}{(\sigma_n^2+g_D^2v_s^2)^{1/2}} \nonumber\\
a_n &=& \frac{\sigma_n \chi_{n}-g_Dv_s V_{5n}}{(\sigma_n^2+g_D^2v_s^2)^{1/2}}\,,
\end{eqnarray}
where $\sigma_n=(n+1/2)/R$ {\footnote {It is occasionally useful to write these expressions in the form employing KK level dependent mixing angles: 
$a_n=\cos \theta_n \chi_n-\sin \theta_n V_{5n}$, \etc, as will be employed below.}}. The Goldstones are, of course, absent in the unitary gauge in which we will work, while the $a_n$ 
KK tower fields acquire physical masses given by\cite{5d}  $(m^a_n)^2=(\frac{n+1/2}{R})^2+g_D^2v_s^2$ and the $a_n$ wavefunctions are given by 
$a_n(y)=\sqrt{\frac{2}{\pi R}} \sin \frac{(n+1/2)y}{R}$. The functional change in the $y$-dependence is a result of the fact that $a_n \sim \partial_y \chi_n$. 
Note the $V_5$ fields do not experience the effects of the gauge BLKT.

For the gauge KK fields we assume that the $v_n$ satisfy the same BCs as in the previous subsection, including the BLKT described by $\delta_A$. However, in the present 
case the KK masses are altered due to the presence of the bulk mass contribution and are now of the form
\begin{equation}
(m^V_n)^2=\Big(\frac{x^V_n}{R}\Big)^2+g_D^2v_s^2\,, 
\end{equation}
where the roots $x^V_n$ are given by the solutions of the equation 
\begin{equation}
\cot \pi x^V_n = \frac{\delta_A}{2x^V_n} \Big[ (x^V_n)^2+(g_Dv_sR)^2\Big]~=\Omega_n\, 
\end{equation}
The corresponding gauge tower wavefunctions are given by an expression very similar to that found in the last section except for a minor but significant difference in the 
normalization factor due to the gauge bulk mass:
\begin{equation}
(N^V_n)^2 = \frac{2}{\pi R}~\Bigg[1+\Omega_n^2+\frac{\delta_A}{\pi}-\frac{\Omega_n}{\pi x^V_n}\Bigg]^{-1}~\Big({\rm sin^2} \pi x^V_n\Big)^{-1}\,,
\end{equation}
which reduces to the previous result in the absence of the bulk gauge mass. Note that these gauge roots are still found to satisfy $x^V_n \leq n+1/2$ for all $n$, 
so that we always have  $m^V_n < m^a_n$. The remainder of the (relative) mass spectrum is set by the size of ratio of couplings $r=2\lambda_S/g_D^2$ which is naturally expected to be 
of $\sim O(1)$ but can be freely chosen.  Note that when $r<1$ the $h_n$ KK states are lighter than the corresponding $a_n$ (which we will assume here) and vice-versa. We also define 
the dimensionless $\sim O(1)$ quantities $h=8\lambda_Sv_s^2R^2$ and $a=4g_D^2v_s^2R^2$, not to be confused with the 5-D fields. Concentrating on the important lightest modes in each 
KK tower we also define the specific mass ratios $m^V_1/m^h_1= \Sigma$ and $m^a_1/m^h_1 = 1+\delta$. These few parameter ratios control the mass spectra, the associated 
couplings and consequently determine the various model signatures. Note that in units of $(2R)^{-2}$, the squared masses of the $h_n,V_n$ and $a_n$ states are given by 
$(2n+1)^2+h, (2x^V_n)^2+a$ and $(2n+1)^2+a$, respectively.

\begin{figure}[htbp]
\centerline{\includegraphics[width=5.0in,angle=90]{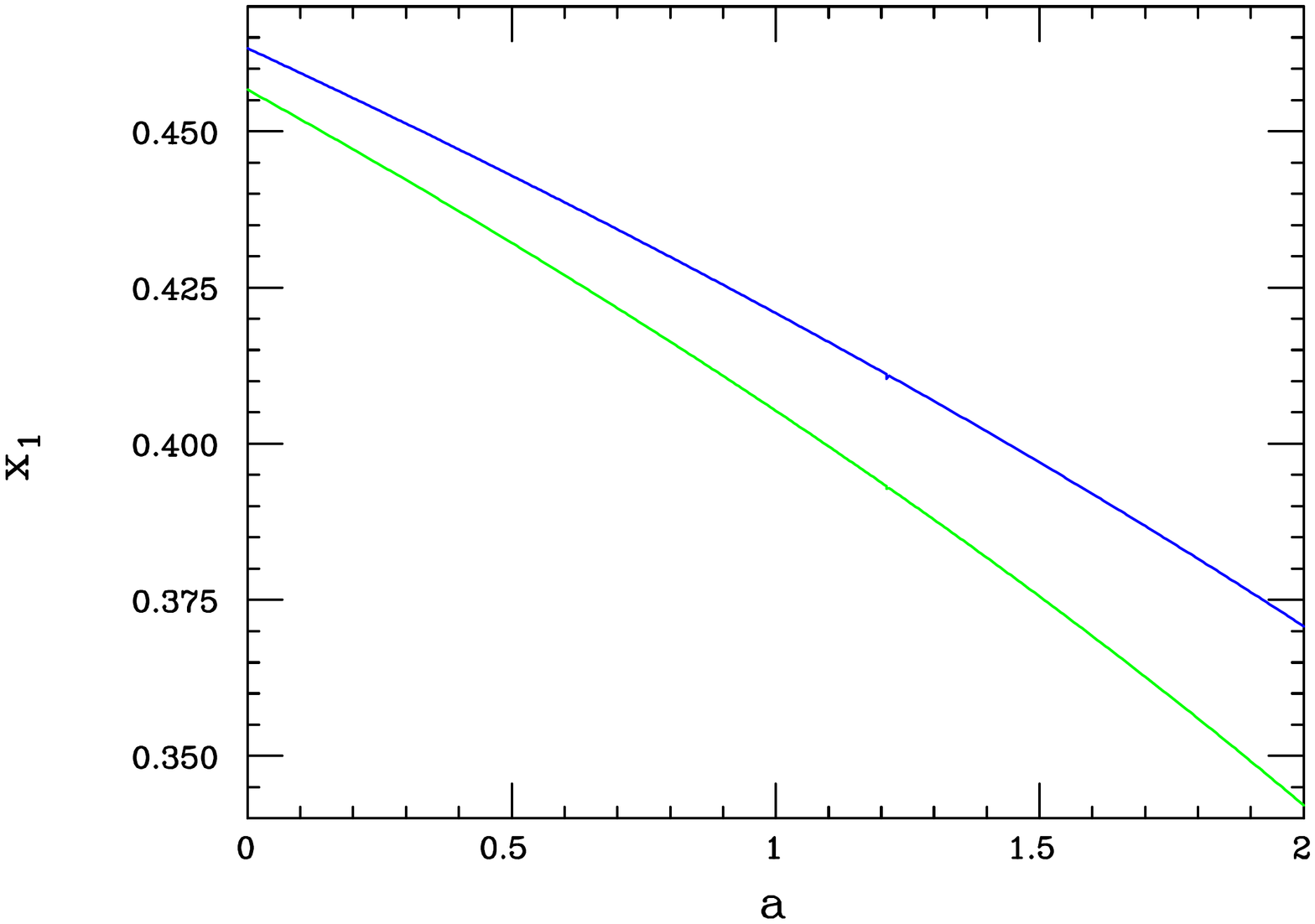}}
\vspace*{-3.1cm}
\centerline{\includegraphics[width=5.0in,angle=90]{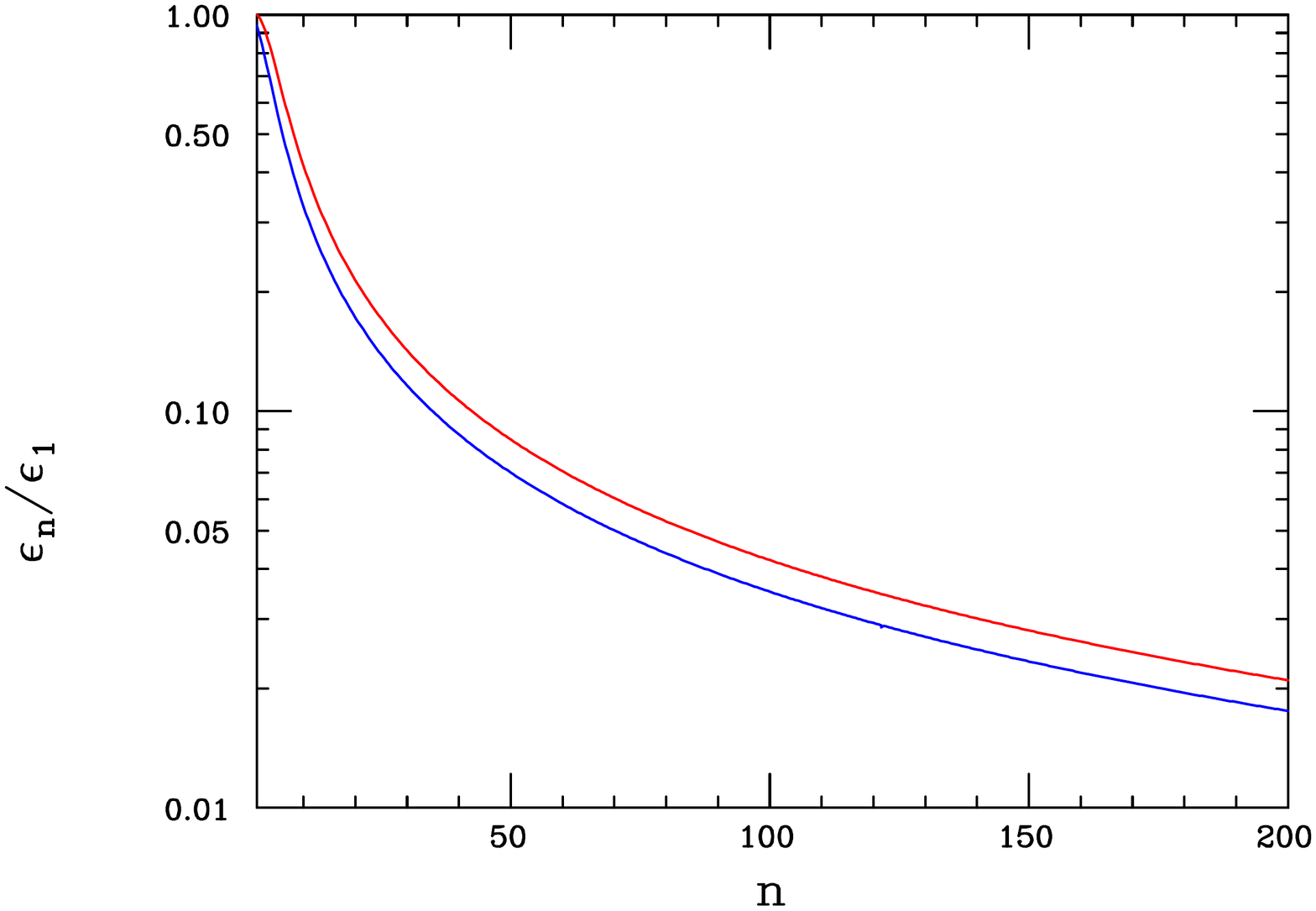}}
\vspace*{-1.90cm}
\caption{Top: Value of $x^V_1$ as a function of $a$ for $\delta_A=0.5$(upper curve) and 0.6(lower curve), respectively. Bottom: Values of $\epsilon_n/\epsilon_1$ as a function of $n$ 
assuming $a=1$ for $\delta_A=0.5$(upper) and 0.6(lower), respectively.}
\label{bm}
\end{figure}

Interesting mass spectra can be found in multiple ways, given that there are now several free parameters to consider, based on the constraints and discussion above; the approach 
we follow is to again choose two interesting benchmark points as in the previous subsection. As we will see, one feature of this set up is that masses associated with the 
3 KK towers will generally be more degenerate than those found in the previous subsection since we are no longer introducing a separate BLKT for the 5-D scalar. Note that since the 
$V_n$ are lighter than the $a_n$, due to the gauge BLKT,  the lightest member of the $h_n$ KK tower, a real scalar field is to be identified with the DM. This implies that we must require 
that $m^h_1<m^V_1$, \ie, $1+h < (2x^V_1)^2+a$, to avoid the CMB constraint when choosing our BM points.  This constraint is powerful in that for fixed values of $a$, $x^V_1$ falls with i
ncreasing $\delta_A$ and for fixed $\delta_A$, $x^V_1$ also falls with increasing $a$. Thus small values of $\delta_A$ and not too large values of $a$ are preferred. The upper panel in 
Fig.~\ref{bm} shows how the numerical value of this root falls with increasing $a$ for two representative values of $\delta_A$. As might be expected, since the $x^V_n$ depend on $a$, 
so will the $\epsilon_n$'s; this is shown in the lower panel of Fig.~\ref{bm}. 

The couplings between the dark scalar and gauge fields are determined by the integrals over the products of the 5-D wavefunctions which takes the form (where the $\cos \theta_m$ mixing 
factor defined above) 
\begin{equation} 
c_{mn}^i = \int_0^{\pi R} ~dy~\cos \theta_m a_m(y)h_n(y)v_i(y)\,
\end{equation}
and, correspondingly, the normalized $\tilde c_{mn}^i$ can also be defined as above. A few obvious differences that we see immediately are:  ($i$) that instead of an $S^\dagger SV$ 
4-D interaction we now have an $haV$ interaction, with a tensor structure of the form $\sim (a\partial_\mu h-h\partial_\mu a)V^\mu$ and ($ii$) the appearance of an overall factor of 
$\cos \theta_m$. This factor occurs because it is actually the weak-basis fields $\chi_m$ which are entering into this interaction and 
so we must project out this part of the $a_m$ fields.  This interaction structure has many important implications for DM physics. First, in DD experiments, at tree-level, the basic 
scattering process is now {\it inelastic} and of the form $h_1 e\to a_1 e$ for scattering off of electrons via DP tower exchange. However, the elastic $h_1e \to h_1 e$ process does occur 
at the1-loop level but is suppressed by the electron mass as well as the usual loop factor leading to an effective operator given by 
\begin{equation} 
{\cal O}= \frac{\alpha}{2\pi}m_eg_D^2\epsilon_1^2~L~\bar eeh_1^2\,
\end{equation}
where $L$ is a product of $\epsilon_i$'s, $\tilde c_{nm}^i$'s and loop functions summed over intermediate states of the $a_n$ and $V_i$'s. (Note that the KK sums here are well-behaved 
since the falling $\epsilon_n$'s suppress the higher KK contributions.) 
This suppression, even with 
$g_D^2\epsilon_1^2 \lsim 10^{-6}$ and $m_{h_1}=25$ MeV, leads to cross sections $\sim 10^{-50}$ cm$^2$ so this process is not likely to be accessible soon. We note that for any 
interesting, non-tuned value of $\delta$, the $a_1-h_1$ mass splitting will always be sufficiently large as to prevent the tree-level inelastic process from occurring. Since $h_1$ is 
lighter than $V_1$ by construction, the DM obtains its thermal relic abundance via co-annihilation, \ie, via the $s-$channel process $h_1a_1\to V_n \to$ SM and so this is again sensitive to
the particular value of $\delta$ (exponentially through the Boltzmann suppression factor). Note that since the $a_1$ is heavier than $V_1$, the process 
$a_1a_1 \to V_1V_1$ via $t,u-$ channel $h_n$ exchange is barely kinematically allowed and is also doubly Boltzmann suppressed. Third, the annihilation of DM {\it today} will likely be 
unobservable, not only due to the p-wave, $v^2$-suppressed cross section but also due to the fact that all of the $a_1$ states needed for co-annihilation will have long since decayed away 
via the 3-body mode $a_1\to h_1V_N^*\to e^+e^-$. The lifetime for this process is also found to be rather sensitive to the value of $\delta$ as expected.  Lastly, while the $a_1a_1$ initial 
state can annihilate into the $h_n$ in the $s$-channel, the $h_n$ themselves do not couple to any of the SM fields 
at tree level since their wavefunctions vanish on the SM brane as discussed above. Lastly, we note that the $V_5$ content of the physical fields cannot 
mediate any interactions with those of the SM.  

We now turn to our specific BMs for this scenario. Since the spectrum is fixed, apart from an overall scale, once we choose our parameters and since the DM mass, $m_{h_1}$, is 
the smallest one we will express all our masses in units of the DM mass below for convenience. For BM1, we will assume $\delta_A=0.5$ as above, $a=1$, $\delta=0.2$ which leads to 
$\Sigma=m_{V_1}/m_{h_1}=1.109$ so that $h=7/18\simeq 0.389$ while for BM2, we take $\delta_A=0.6$, $a=1$, $\delta=0.14$, $h=0.4$ and thus $\Sigma=m_{V_1}/m_{h_1}=1.096$, 
respectively. Tables~\ref{spec2} and ~\ref{spec3} display the masses of the lowest lying KK states for these two BM models. Once the values of 
$\delta_A$ and $a$ are fixed all of the couplings $\tilde c_{mn}^i$ are also fixed and these {\it do not} depend upon the numerical values of $h,\delta$. 
We note that we can switch the parameter ordering in generating further BM points. Choosing a pair of $\Sigma,\delta$ values, $h$ 
becomes fixed and we can determine the value of the lowest gauge KK root as a function of $a$. Using the equation for the gauge roots then gives us $\delta_A$. For example, 
with $\Sigma=1.1,\delta=0.2$ and $a=1$, we obtain $\delta_A\simeq 0.554$, $x^V_1\simeq 0.4125$ and $h=7/18$.

\begin{table}
\centering
\begin{tabular}{|l|c|c|c|} \hline\hline
KK level&     V      &  h    &   a     \\
\hline  
~~~~1   &   1.109     &   1.000   &   1.200    \\
~~~~2   &   2.493     &   2.600   &   2.683     \\
~~~~3   &   4.038     &   4.276   &   4.327    \\
~~~~4   &   5.627     &     -    &    -      \\
\hline\hline
\end{tabular}
\caption{Masses of the lightest gauge and scalar KK states in units of $m_{h_1}$ for the BM1 point considered in the text.}
\label{spec2}
\end{table}
\begin{table}
\centering
\begin{tabular}{|l|c|c|c|} \hline\hline
KK level&     V      &  h    &   a     \\
\hline  
~~~~1   &   1.036     &   1.000   &   1.150    \\
~~~~2   &   2.359     &   2.508   &    2.571    \\
~~~~3   &   3.829     &   4.107   &    4.146    \\
~~~~4   &   5.348     &     -    &    -      \\
\hline\hline
\end{tabular}
\caption{Masses of the lightest gauge and scalar KK states in units of $m_{h_1}$ for the BM2 point considered in the text.}
\label{spec3}
\end{table}

\begin{figure}[htbp]
\centerline{\includegraphics[width=5.0in,angle=90]{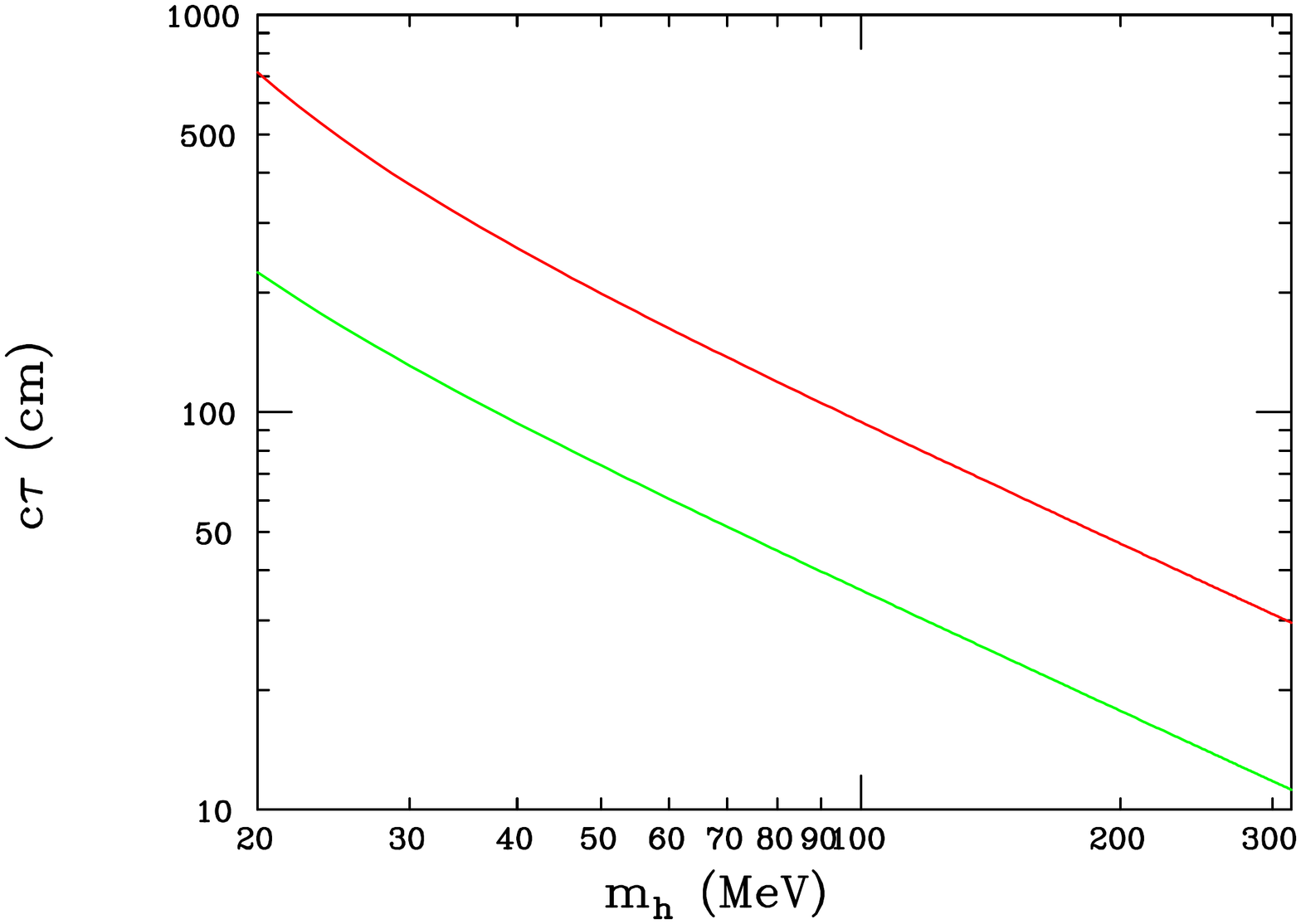}}
\vspace*{-3.1cm}
\centerline{\includegraphics[width=5.0in,angle=90]{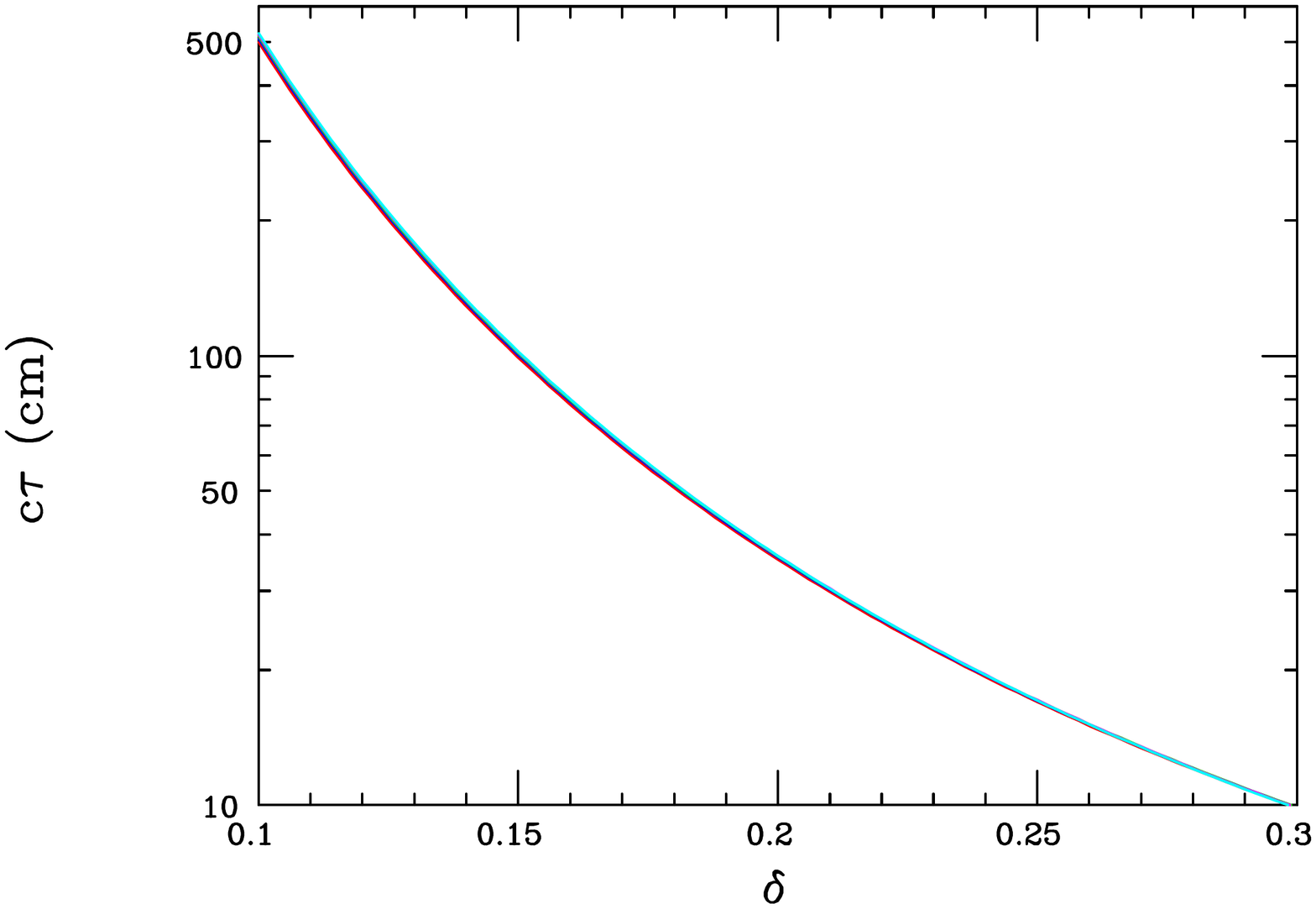}}
\vspace*{-1.90cm}
\caption{Top: Unboosted decay lengths for the $a_1$ in BM1(lower curve) and BM2(upper curve) as functions of $m_{h_1}$. Bottom: $\delta$-dependence of the $a_1$ decay length 
for $m_{h_1}=100$ MeV, $\delta_A=1$ for different choices of $h=0.1-0.5$, which essentially lie atop one another. Note that in both panels $g_D\epsilon_1=10^{-3}$ has been assumed.}
\label{life}
\end{figure}

As in the previous subsection, $V_1$ can decay only to the SM, most likely $e^+e^-$ pairs. However, as noted, $a_1(m_a)$ can decay 
to $h_1(m_h)$ via an off-shell $V_i(m_{V_i})$ KK tower, materializing as $e^+e^-$ as well since this mass splitting is likely to be greater than 1 MeV. The partial width for this 3-body 
process employing\cite{pdg} the traditional variables $m_{ij}^2=(p_i+p_j)^2$ is given by
\begin{equation} 
\Gamma_3=\frac{(eg_D\epsilon_1)^2}{128\pi^3m_a^3} \int  dm_{12}^2 \int dm_{23}^2 ~S\Bigg[ (m_a^2-m_h^2)^2-(m_{23}^2-m_{13}^2)^2 +m_{12}^2
\Big(m_{12}^2-2(m_h^2+m_a^2)\Big)\Bigg]\,
\end{equation}
where $S$ symbolizes the KK sum (away from resonances) given by 
\begin{equation} 
S=\sum_{ij} (-1)^{i+j} \frac{(\epsilon_i\epsilon_j/\epsilon_1^2) \tilde c_{11}^i \tilde c_{11}^j}{(m_{12}^2-m_{V_i}^2)(m_{12}^2-m_{V_j}^2)}\,
\end{equation}
Here, $4m_e^2\leq m_{12}^2\leq (m_a-m_h)^2$ is the square of the $e^+e^-$ pair mass, \etc, and $m_{12}^2+m_{13}^2+m_{23}^2=2m_e^2+m_a^2+m_h^2$ with  
$p_{1,2,3}$ being identified with the momenta of $e^\pm, a$, respectively. The top panel of Fig.~\ref{life} shows the resulting unboosted decay lengths of the $a_1$ in our two BM 
scenarios as a function of the $h_1$ mass. Here we see that the typical $a_1$ decay length lies in the $\sim10-1000$ cm range but this scales as 
$(10^{-3}/g_D\epsilon_1)^2$ for other choices of these parameters.  To get a feel for the $\delta$-dependence of these decay lengths, we fix $m_{h_1}=100$ MeV, $\delta_A=1$ and 
take various choices of $h=0.1-0.5$ in the lower panel and determine $c\tau$ vs $\delta$; note that the $h$-dependence itself is rather weak. In this parameter range $c\tau$ is found 
to scale roughly as $\sim \delta^{-3.6}$. 

The heavier $h,a$ and $V$ KK states decay in a manner similar to that in the previous subsection except that the existence of three distinct towers adds complexity. However, given that 
the $haV$ vertex is the only one relevant for any gauge mediated two-body decays, these are necessarily of the form $V\to ah$, $a\to Vh$, $h\to Va$, \etc. The expressions 
for the various partial widths are similar to those encountered in the last subsection. A short consideration tells us that the BFs for the processes $V_2\to h_1a_1$,  $h_2\to V_1a_1$ 
and $a_2\to V_1h_1$ are all $\sim$100\% for both BMs. The corresponding BFs for the next set(s) of KK states can be found in Table~\ref{decay2}. Here we see that for many of the modes 
there are no large differences between the two BMs; this is a result of the rather similar, compressed spectra in both cases. The most significant differences are visible in the decay modes 
with relatively small BFs for the higher gauge KK states. In this setup we note that when the heavier gauge KK states are directly produced from SM fields they will cascade 
through various channels down to (potentially numerous) combinations of the three $h_1,a_1$ and $V_1$ fields, similar to what we described in the last subsection. While the $h_1$ will produce 
missing energy and $V_1$ will decay (possibly with a long lifetime) into an $e^+e^-$ pair, the $a_1$ will produce {\it both} missing energy and a soft $e^+e^-$ pair after traveling $\sim 1$ m. 
These final states would be interesting to observe. Unlike in the scenario discussed in the previous subsection, if $V_1$ {\it can} decay to pairs of DM particles, 
long-lived $a_1$'s will still be produced in the cascade decays of the more massive states. In such a case, missing energy plus multiple displaced $e^+e^-$ production vertices will yield a 
rather clean signature.

\begin{table}
\centering
\begin{tabular}{|l|c|c|c|} \hline\hline
~~~Process             &    BF(BM1)    &   BF(BM2)   \\
\hline
$h_3\to V_1a_1$ &  44.27       &      44.03       \\
$h_3\to V_1a_2$ &  34.37       &      34.51      \\
$h_3\to V_2a_1$ &  21.36       &      21.45     \\
                     &        &    \\
$a_3\to V_1h_1$ &   14.88      &      14.91       \\
$a_3\to V_2h_1$ &   23.84      &      23.37       \\
$a_3\to V_1h_2$ &   61.28      &      61.72      \\
      &          &    \\ 
$V_3\to a_1h_1$ &   23.54      &      29.72       \\
$V_3\to a_2h_1$ &   37.14      &      34.80       \\
$V_3\to a_1h_2$ &   39.33      &      35.48       \\
       &         &        \\
$V_4\to a_1h_1$ &   2.99        &      3.36       \\
$V_4\to a_2h_1$ &   8.85        &      9.99       \\
$V_4\to a_3h_1$ &   5.72        &      3.19      \\
$V_4\to a_1h_2$ &   20.83       &     23.21      \\
$V_4\to a_2h_2$ &   55.64       &     57.11        \\
$V_4\to a_1h_3$ &    5.97        &     3.15      \\
\hline\hline
\end{tabular}
\caption{Branching fractions for the various decay modes in per cent for the next highest gauge and $h,a$ KK states in both BM scenarios as discussed in the text.}
\label{decay2}
\end{table}

\begin{figure}[htbp]
\centerline{\includegraphics[width=5.0in,angle=90]{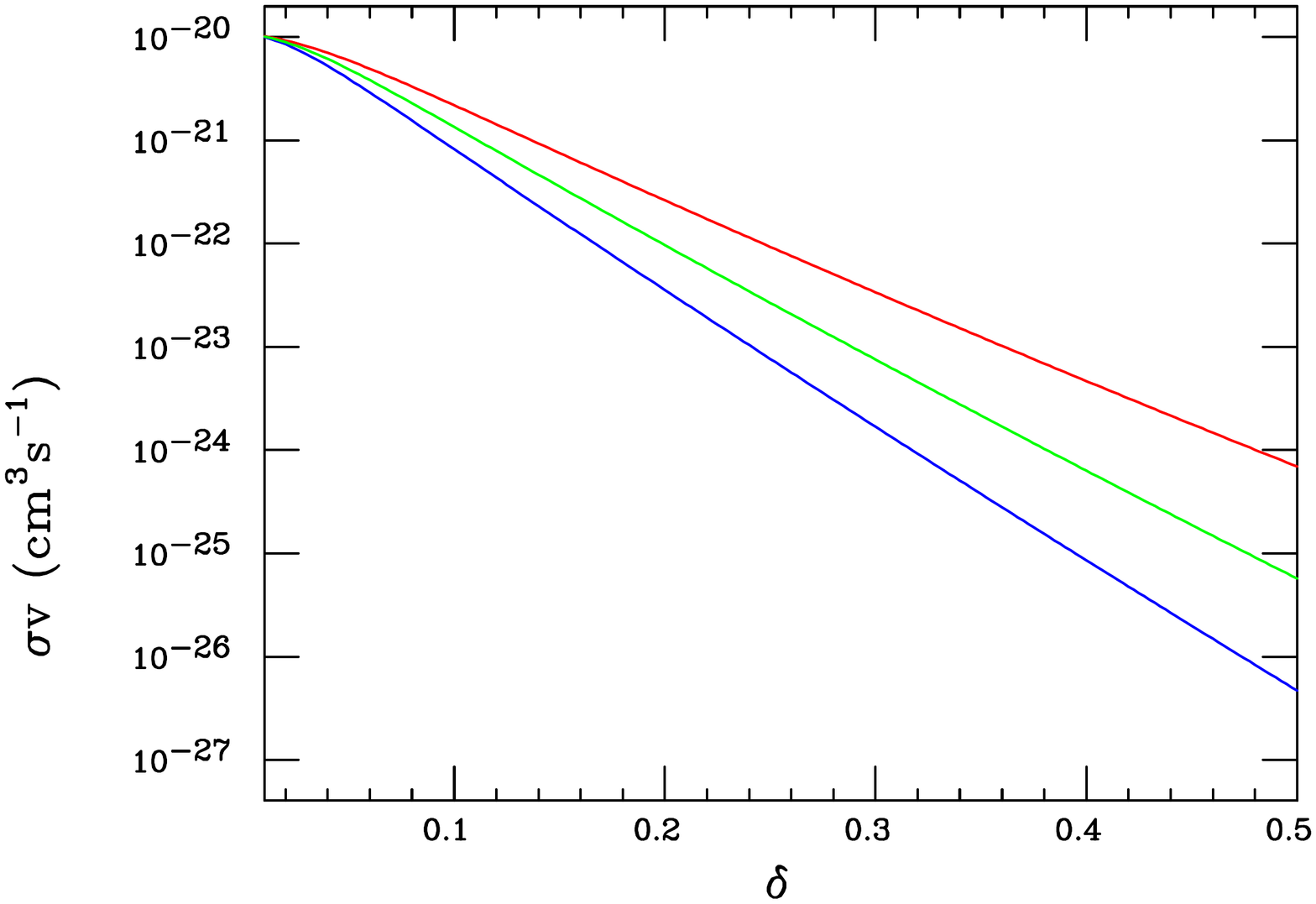}}
\vspace*{-3.1cm}
\centerline{\includegraphics[width=5.0in,angle=90]{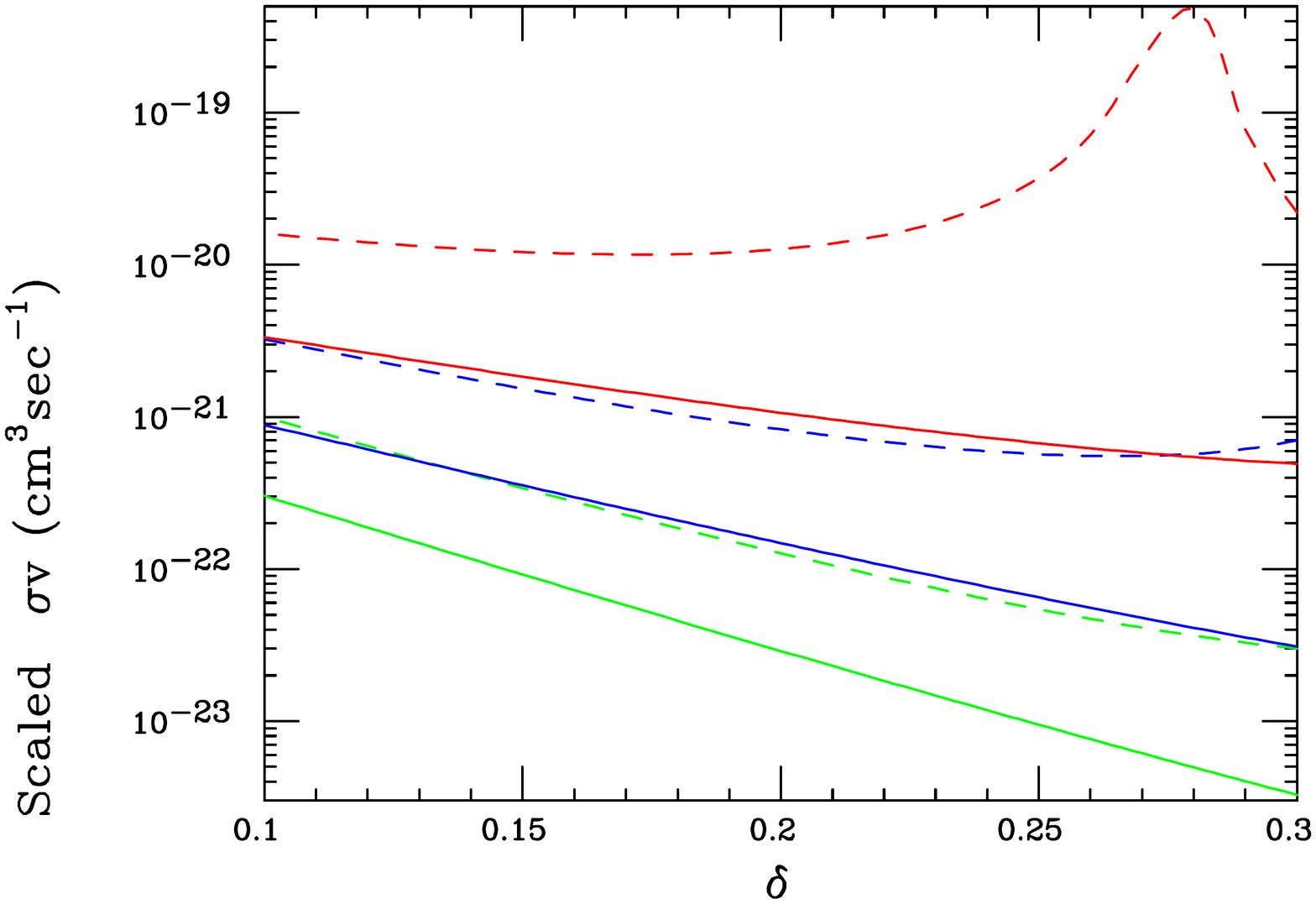}}
\vspace*{-1.90cm}
\caption{Top: Explicit dependence of the annihilation cross section on the parameter $\delta$ with all others held fixed demonstrating the effects of Boltzmann suppression. From top to
bottom the curves are for $x_F=20(25,30)$, respectively. Bottom: Examples of thermally weighted annihilations cross sections as functions of $\delta$ assuming  $\delta_A=0.5$; the 
red(blue,green) curves correspond to $x_F=20(25,30)$ while the solid(dashed) correspond to $h=0.3(0.4)$, respectively. }
\label{stuff1}
\end{figure}

We now turn our attention to the DM relic density; the relevant process for this is $h_1(m_h)a_1(m_a) \to V_i^*(m_{V_i}) \to e^+e^-$, plus other possible SM final 
states if phase space is available; here we limit ourselves to the $e^+e^-$ mode. In the $m_e\to 0$ limit we obtain, away from the narrow $V_{i,j}$ resonance peaks: 
\begin{equation} 
\sigma v=\frac{(g_De\epsilon_1)^2 s}{12\pi}\Big(1-\frac{(m_a+m_h)^2}{s}\Big)\Big(1-\frac{(m_a-m_h)^2}{s}\Big)~\sum_{i,j} (-1)^{i+j} \frac{(\epsilon_i\epsilon_j/\epsilon_1^2) 
\tilde c_{11}^i \tilde c_{11}^j}{(s-m_{V_i}^2)(s-m_{V_j}^2)}\,
\end{equation}
There are two interesting aspects of this DM co-annihilation process\cite{Davoli:2017swj}, one being the mass splitting between the $a_1$ and $h_1$, controlled by 
$\delta$, that leads to significant Boltzmann suppression of the annihilation rate. As is well-known\cite{Feng:2017drg}, this modifies the DM co-annihilation rate by a factor of
\begin{equation} 
\sigma v_{eff}= \frac{2\alpha}{1+\alpha^2} ~ \sigma v_{ha}\,
\end{equation}
where $\sigma v_{ha}$ is the the $h_1-a_1$ annihilation cross section above and $\alpha =(1+\delta)^{3/2}e^{-x\delta}$ with $x=m_{h_1}/T\simeq x_F\sim 20-30$. If this was the {\it sole} 
effect, the annihilation cross section would fall off drastically with increasing $\delta$ as can be seen in the top panel of Fig.~\ref{stuff1} where all other quantities are held fixed and we  
vary $\delta$ for three specific values of $x_F$. 
\begin{figure}[htbp]
\centerline{\includegraphics[width=5.0in,angle=90]{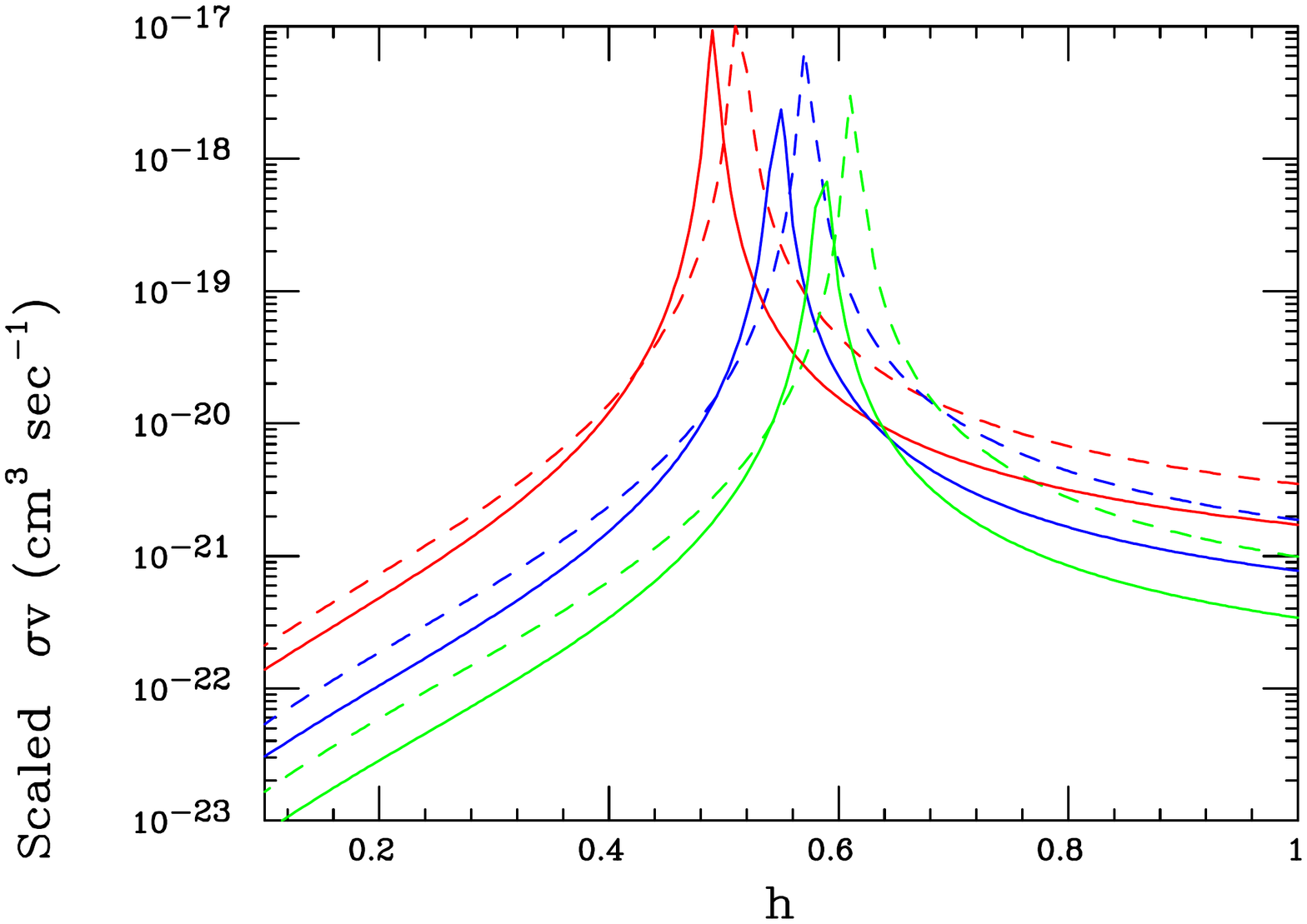}}
\vspace*{-3.1cm}
\centerline{\includegraphics[width=5.0in,angle=90]{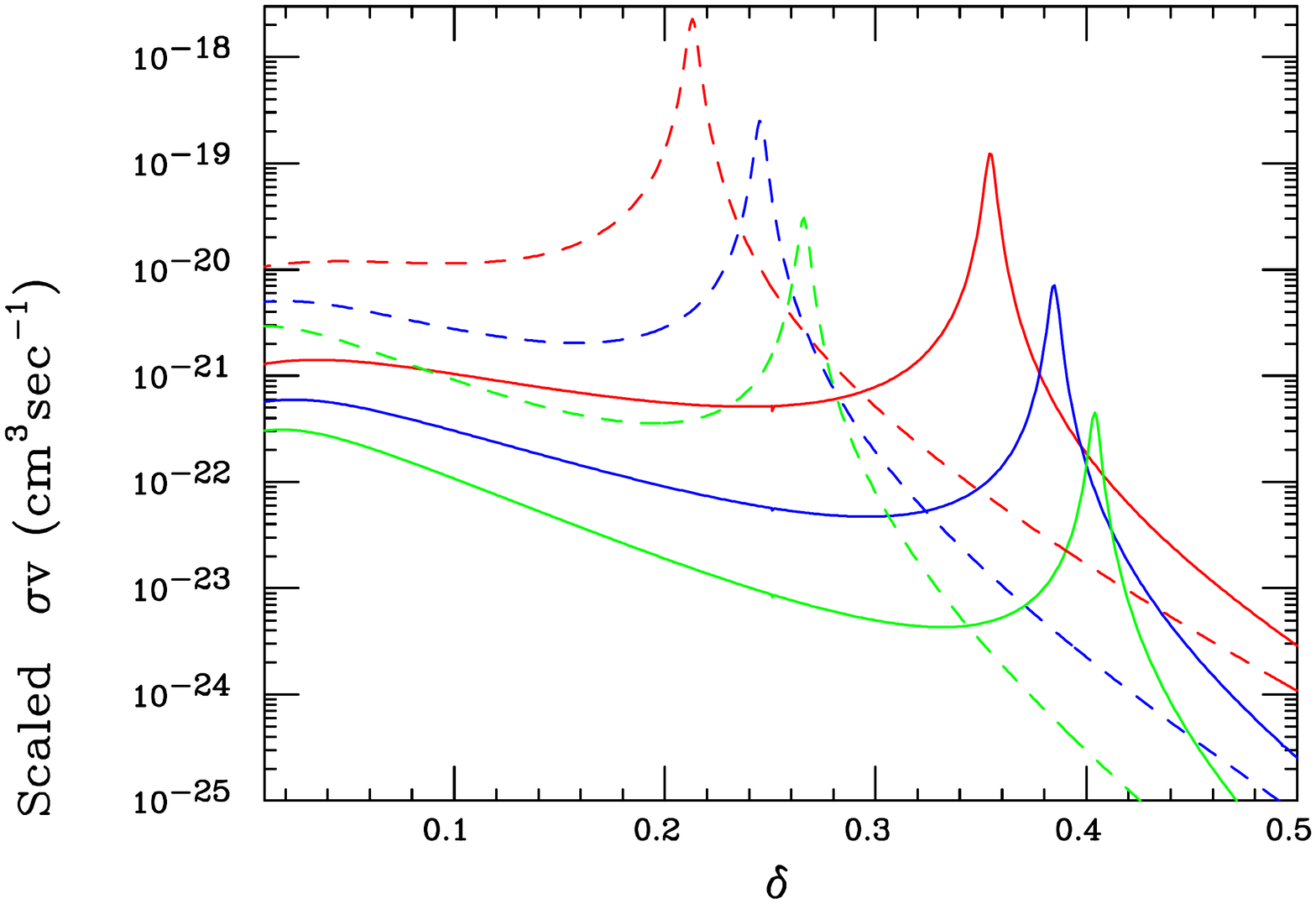}}
\vspace*{-1.90cm}
\caption{Scaled annihilation cross sections for BM1(solid) and BM2(dashed) for $x_F=$20(red), 25(blue) and 30(green) as functions of $h$(top) or $\delta$(bottom) with all other BM 
parameters held fixed. To get the true cross section, these results need to be rescaled by a factor of $(g_D\epsilon_1/m_{DM}(\rm {GeV}))^2$. }
\label{stuff2}
\end{figure}
The second aspect of relevance is the proximity of the sum of the initial state masses, $m_{a_1}+m_{h_1}$, to that of the {\it next} lowest lying gauge KK mass, 
$m_{V_2}$.{\footnote {A similar situation can arise in the case of UED\cite {ued2}.}}  By construction, the sum of the $a_1,h_1$ masses lies above that of $V_1$; however, we note that 
this sum is, for both BMs, not far below that of $V_2$ implying 
that partial resonant enhancement may play a role in off-setting the Boltzmann suppression. This is particularly true as the thermal velocities of $h_1,a_1$ can raise 
the center of mass energy in the collision process closer to the $V_2$ resonance region. To probe the potential importance of the $V_2$ resonance, we examine the scaled 
annihilation cross sections for the BM models at fixed temperature (\ie, fixed $x_F$) and with the corresponding fixed values of the velocities but allowing for one of the parameters 
$h,\delta$ to differ from their chosen BM values; this is shown in Figs.~\ref{stuff1} and ~\ref{stuff2}. 
While varying the value of $h$ simply moves the $h_1,a_1$ spectrum for fixed $\delta$ so that the $V_2$ resonance is encountered, raising $\delta$ both lets us encounter this 
resonance while also simultaneously suppressing the overall cross section. Note that for the true BM parameters, taking $(g_D\epsilon_1/m_{DM}(\rm {GeV}))^2 \sim 10^{-5}$ would roughly 
yield the desired value of the relic density. Thus the nearby resonance can, at least locally, appear to compensate for the overall Boltzmann suppression. 
\begin{figure}[htbp]
\vspace*{-1.5cm}
\hspace*{-0.1cm}
\centerline{\includegraphics[width=5.0in,angle=90]{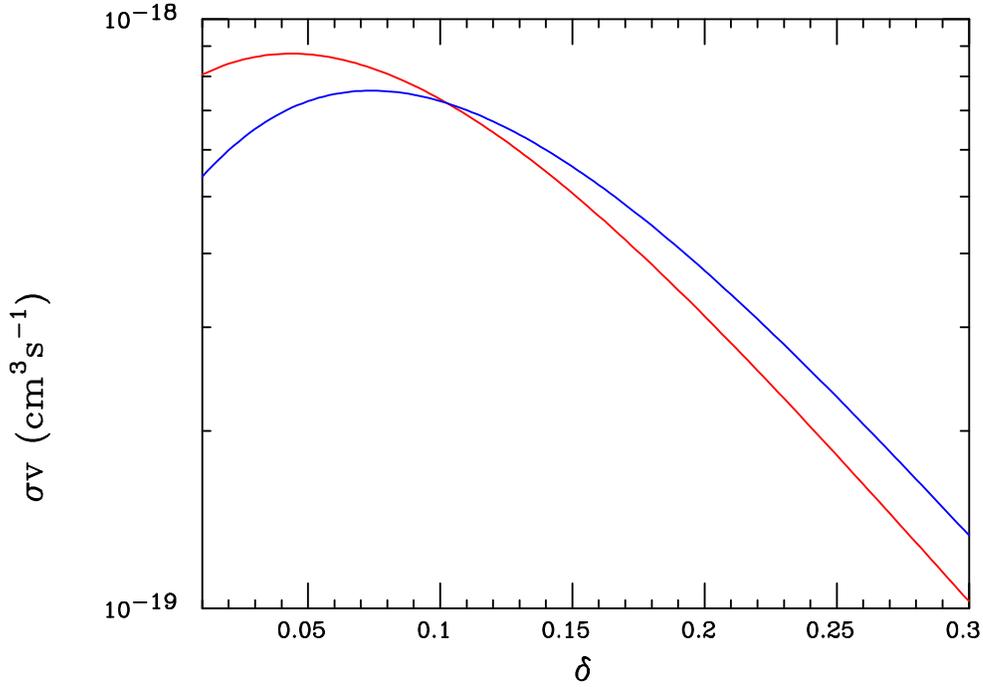}}
\vspace*{-1.50cm}
\caption{Thermally averaged $h_1a_1\to e^+e^-$ annihilation cross sections for the BM1(red) and BM2(blue) points generalized to functions of the parameter $\delta$ in units of 
$(g_D \epsilon_1/m_{h_1}(GeV))^2$.} 
\label{vweight}
\end{figure}

In order to full understand the combined impact of both the resonances and Boltzmann suppression acting simultaneously on the thermally averaged cross section we return to the 
basic formalism and compute (numerically) the velocity-weighted cross section (with $x_F=20$ assumed here): 
\begin{equation} 
<\sigma v>= \frac{\int d^3p_a \int d^3p_h e^{-E_a/T}e^{-E_h/T} ~ \sigma v_{ha}}{\int d^3p_a \int d^3p_h e^{-E_a/T}e^{-E_h/t}}\,
\end{equation}
with $E_{a,h}$ being the incoming, velocity-dependent energies of $a_1,h_1$ in the collision process, respectively. Note that BM2 has a 
slightly smaller value of $\delta$, 0.12 vs. 0.15,  so that the Boltzmann suppression is reduced and also the sum of the $a_1,h_1$ masses is closer to that of $V_2$ to that resonance effects 
should also be larger. However, since the $\epsilon$'s are somewhat smaller for BM2 since $\delta_A$ is greater, there is for BM2 a small suppression of the couplings 
of the gauge KK states above the lightest mode relative to those of BM1. Explicitly for BM1(BM2) we obtain the cross sections 
$<\sigma v>= 2.5(2.8)\cdot 10^{-26} {\rm {cm^3 s^{-1}}} (g_D\epsilon_1/10^{-8})^2 ~(500 \rm MeV/m_{h_1})^2$.  Here we see that the reduced Boltzmann suppression and resonance 
proximity can compensate for each other and the reduced KK gauge couplings and that reasonable parameter choices can lead to the observed relic density for both BMs. These results are 
shown more generally as functions of $\delta$ in Fig.~\ref{vweight}. Here we see in particular that the resonance is sufficiently Doppler broadened to increase the cross sections for these BMs 
by roughly an order of magnitude or more compared to naive expectations.

\subsection{Model 2':  The DM Lifetime Problem}

As mentioned above, the setup in the previous subsection has an important flaw in that the state $h_1$ can decay to SM fields (two pairs of $e^+e^-$) via double off-shell $V$ tower  
exchange and this decay happens too rapidly for $h_1$ to be the DM. However, there is an easy solution to this problem which leaves the previously discussed phenomenology 
intact with only very minor numerical changes. The first step is to make $a_1$ the potential DM state and to do that we must decrease the $a_1,h_1$ masses relative to $V_1$; this 
can be accomplished by introducing a scalar BLKT, $\delta_S> \delta_A$, as was done in an earlier subsection above. Next, by requiring $r>1$, we make $h_1$ heavier than $a_1$ 
while simultaneously maintaining the constraint $m_{h_1}+m_{a_1}>m_{V_1}$. This essentially just interchanges the roles of $a_1$ and $h_1$ in the previous section without any 
other phenomenological impact. Of course, $a_1$ remains unstable but, since it is CP-odd, it decays via a {\it triple} off-shell decay chain yielding a 6-body final state: 
$a_1 \to V^*h_1^*,~h_1^*\to V^*V^*$ followed by $V^* \to e^+e^-$.  A very rough estimate of this partial decay width given by  
\begin{equation} 
\frac{\Gamma_6}{m_{a_1}} \sim\frac{\alpha_D^3 \alpha_{em}^3 \epsilon_1^6}{(3!)^2 ~(2\pi)^{14}}~ \frac{2\pi^8}{2^6\Gamma(8)}~\Big(\frac{m_{a_1}}{m_{V_1}}\Big)^{12}\,,
\end{equation}
which follows from the coupling constants, six-particle final state phase space factors, the appearance of two triplets of identical particles in the final state, the assumption that 
$h_1$ is only slightly more massive than the $a_1$, and the dominance of $V_1$ in the off-shell gauge boson propagator summations.  
Taking $\alpha_D \simeq \alpha_{em}$ and $\epsilon_1=10^{-4}$ as above, $m_{a_1}/m_{V_1}=1/2$, and $m_{a_1}=50$ MeV one obtains 
$\Gamma_6 \sim  3 \cdot 10^{-51}$ MeV corresponding to a lifetime of $\sim 2 \cdot 10^{29}s$ which is a factor of $\sim 10^4$ longer than the lower bound from the 
CMB\cite{Slatyer:2016qyl} and $\sim 100$ times longer than the tentative new limits from the 21 cm line\cite{21club} for this mass. Clearly a more detailed calculation of this partial 
width is warranted but this rough estimate indicates that the $a_1$ is very likely to be sufficiently long-lived in this setup to act as the DM without much impact on the analysis 
presented in the previous subsection.

\section{Discussion and Conclusions}

In this paper we have considered an extension of the familiar 4-D kinetic-mixing portal/dark photon model to 5-D by adding a single, flat extra dimension of inverse `radius' 
$R^{-1}\sim 10-1000$ MeV, which is treated as an interval allowing for more general boundary conditions. In the simple models we construct, while the gauge mediator and 
the dark matter experience the full 5-D, the SM fields are localized to a 4-D brane at one end of this interval. To avoid constraints from CMB measurements we have considered 
the case of complex scalar dark matter with the lightest scalar Kaluza-Klein state lying below that of the corresponding lightest dark photon KK state to insure p-wave annihilation. 
In these models, the consistency of the field redefinitions needed to undo the effects of KM and the avoidance of ghosts and/or tachyons requires the 
existence of a BLKT on the brane where the 4-D SM fields are localized. We constructed two distinct scenarios depending upon whether or not the scalar dark matter field 
obtained a vev in the 5-D bulk. The presence of this vev was shown not to be required in order to break the dark $U(1)_D$ gauge symmetry as this was accomplished by the 
choice of BCs with the fifth component of the gauge fields then acting as the Goldstone bosons.  Constraints arising from precision electroweak measurements, from rare $Z$ 
and Higgs decays as well as from monojet searches at the LHC were shown to be satisfied even in the presence of multiple KK excitations. We also found that we can 
use these BCs to remove the possibility of a Higgs portal that induces a coupling of the SM Higgs with the KK tower scalar fields thus avoiding potential conflicts 
with constraints on exotic Higgs decays. If the bulk scalar does not obtain a vev, the dark matter remains as a complex field whereas when the vev is present the complex 
scalar decomposes into CP-even and CP-odd scalar towers. When the vev is absent the required ordering of the scalar vs gauge mass spectrum necessitates the addition of 
a BLKT for the scalar field on the non-SM brane. In the setup where a vev occurs the CP-even scalars are mass eigenstates while the CP-odd field mixes with the 
fifth component of the gauge field to form both Goldstone KK and physical CP-odd scalar towers. Without a scalar BLKT, level-by-level these CP-odd scalar fields 
were determined to be more massive than the corresponding gauge states due to the gauge BLKT; a BLKT for the scalar can change this hierarchy. Indirect detection of dark 
matter today in either scenario was shown to be unlikely: in the first setup, due to the p-wave nature of the annihilation process, the cross section is suppressed due to the low 
velocity of the dark matter. In the second scenario, the relic density is achieved via p-wave co-annihilation with the opposite CP scalar. In the first scenario, a small but potentially 
observable direct detection, spin-independent scattering cross section off of bound atomic electrons was found whereas in the second such a cross section was determined to be 
unobservably small due to loop suppression with the corresponding tree-level, now inelastic, scattering process being kinematically forbidden for non-relativistic dark matter velocities. 

\begin{figure}[htbp]
\vspace*{-1.5cm}
\hspace*{-0.1cm}
\centerline{\includegraphics[width=5.0in,angle=90]{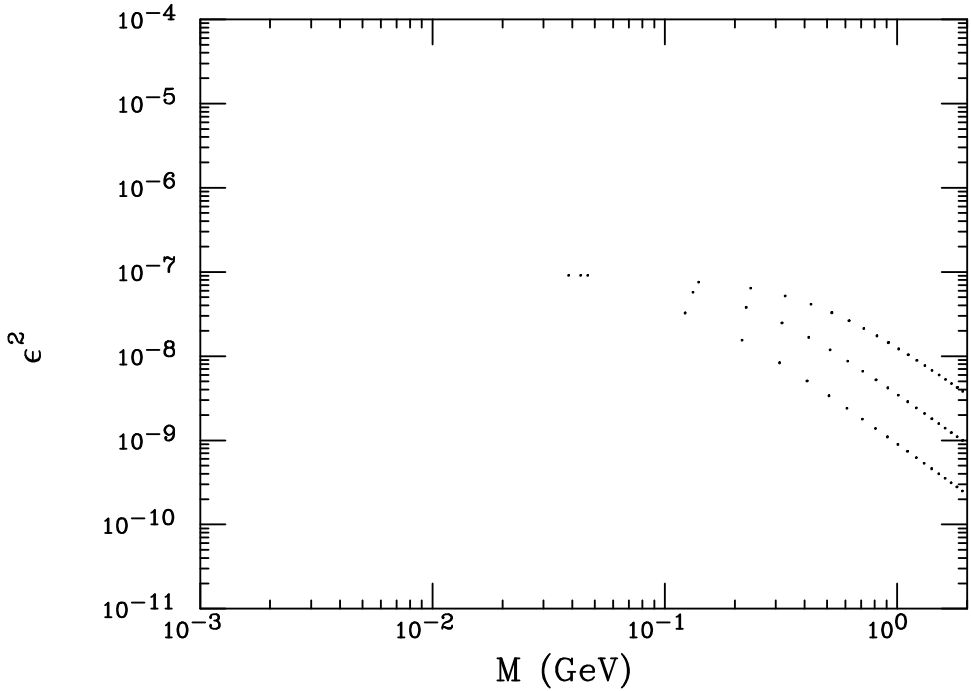}}
\vspace*{-1.50cm}
\caption{Sample loci of points for Model 1 assuming $R^{-1}=100$ MeV and $\epsilon_1=3\cdot 10^{-4}$ for, from top to bottom, $\delta_A=0.1,1,2$, respectively.}
\label{nplot}
\end{figure}

The overall allowed parameter space of the pair of models that we have constructed is somewhat difficult to visualize. In attempting this and to be as model-independent as possible 
is likely best to focus on the DP KK masses and their couplings to SM matter as represented by the $\epsilon_n$.  Semi-quantitatively, the simplest picture to keep in mind is to 
imagine the oft-shown\cite{vectorportal} $\epsilon^2-m_V$ mass plane as this occurs in all the types of models that we consider; see, \eg, Fig.~22 in  Ref.\cite{Battaglieri:2017aum}. 
Once an (experimentally allowed) pair of values for $\epsilon_1,m_{V_1}$ is chosen, the model can be very crudely represented by this point  plus the (asymptotically correct) locus 
of all the subsequent set of pairs of points $m_{V_n}\simeq (2n-1) m_{V_1}$ and $\epsilon_n \simeq \epsilon_1/(n\delta_A)$  in this plane.  Detailed models will of course differ from 
this crude picture by $O(1)$ effects but this image roughly captures the main features of the allowed  parameter space. To give a specific example and provide a clearer understanding 
of this, consider Model 1 above taking $R^{-1}=100$ MeV, $\epsilon_1=3\cdot 10^{-4}$ and $\delta_A=0.5, 1, 2$, which are values typical of the previous discussions above. Fig.~\ref{nplot} 
then shows the loci of these KK points in the $\epsilon^2-m_V$ plane that can be directly compared with Fig.~22 in Ref.\cite{Battaglieri:2017aum}. Presently, for these chosen parameter 
values, the KK's are seen to lie in an allowed region between experimentally excluded areas. Clearly, we could just as easily have chosen an different $R^{-1},\epsilon_1$ `origin' point 
for this loci but the overall shape of the predicted parameter space would be reasonably similar. It is important to remember, however, that these KK states will generally not decay in the 
same manner as the 4-D states of the same mass as represented in Ref.\cite{Battaglieri:2017aum}.

A unique feature of these 5-D models is the simultaneous production of several members of the dark photon tower each of which cascade decays to pairs of 
kinematically accessible scalar tower states. As in the 4-D case several production mechanisms are possible with the final signals being determined by the relative values of the 
masses of the lightest KK tower states.  When the DM is less than half of the mass of the lightest gauge state, this cascade results in a missing energy signature. In associated 
production experiments, such as in meson decays or in inclusive $e^+e^- \to \gamma +X$ reactions, the presence of the individual KK tower states may be reconstructed. In electron 
fixed-target scattering experiments where only the outgoing electron is measured, on the other hand, only the cumulative effect of production and decay of the gauge KK towers is 
observable which may in many cases be difficult to differentiate from the conventional 4-D signal depending on the detailed parameters values. Perhaps even more interesting is the 
situation when the DM is somewhat heavier than the lightest gauge KK state; in such cases the cascades not only produce missing energy but also multiple pairs of light 
charged SM fields, \eg, electrons and/or muons. However, when the DM scalar obtains a vev, then numerous long-lived states are also automatically produced as part of these 
cascades. In either case these are rather striking and distinctive experimental signatures for the 5-D scenario.

The theoretical landscape of DM models continues to broaden and the experimental searches must continue to widen if we are eventually going to corner DM.

\section*{Acknowledgements}

The author would like to thank J.L. Hewett, D. Rueter and G. Wojcik for valuable discussions. The author would like to thank D. Morrissey for bringing  the related work in 
Ref.~\cite{morrissey} to his attention after the first version of the present paper appeared. The author would also like to thank T. Slatyer for discussions on the CMB  constraints for 
light dark matter masses. This work was supported by the Department of Energy, Contract DE-AC02-76SF00515.




\begin{thebibliography}{99}

\bibitem{Arcadi:2017kky} 
  For a recent review of WIMPs, see G.~Arcadi, M.~Dutra, P.~Ghosh, M.~Lindner, Y.~Mambrini, M.~Pierre, S.~Profumo and F.~S.~Queiroz,
  arXiv:1703.07364 [hep-ph].
 
\bibitem{Kawasaki:2013ae} 
  M.~Kawasaki and K.~Nakayama,
  Ann.\ Rev.\ Nucl.\ Part.\ Sci.\  {\bf 63}, 69 (2013)
  [arXiv:1301.1123 [hep-ph]].

\bibitem{Graham:2015ouw} 
  P.~W.~Graham, I.~G.~Irastorza, S.~K.~Lamoreaux, A.~Lindner and K.~A.~van Bibber,
  Ann.\ Rev.\ Nucl.\ Part.\ Sci.\  {\bf 65}, 485 (2015)
  [arXiv:1602.00039 [hep-ex]].

\bibitem{susy17}
For very recent summaries of the dark matter searches at the LHC, in indirect detection and in direct detection, see the talks by H. Flaecher, R. Mahapatra and C. Weniger, respectively,  
given at the {\it 25th International Conference of Supersymmetry and the Unification of the Fundamental Interactions(SUSY17)}, 11-15 Dec 2017, TIFR, Mumbai, India.

\bibitem{Alexander:2016aln} 
  J.~Alexander {\it et al.},
  arXiv:1608.08632 [hep-ph].

\bibitem{Battaglieri:2017aum} 
  M.~Battaglieri {\it et al.},
  arXiv:1707.04591 [hep-ph].

\bibitem{ED}
See, for example, 
  I.~Antoniadis,
  Phys.\ Lett.\ B {\bf 246}, 377 (1990);
  K.~R.~Dienes, E.~Dudas and T.~Gherghetta,
  Phys.\ Lett.\ B {\bf 436}, 55 (1998)
  [hep-ph/9803466] and 
  Phys.\ Lett.\ B {\bf 436}, 55 (1998)
  [hep-ph/9803466];
  I.~Antoniadis, N.~Arkani-Hamed, S.~Dimopoulos and G.~R.~Dvali,
  Phys.\ Lett.\ B {\bf 436}, 257 (1998)
  [hep-ph/9804398];
  N.~Arkani-Hamed, S.~Dimopoulos and G.~R.~Dvali,
  Phys.\ Lett.\ B {\bf 429}, 263 (1998)
  [hep-ph/9803315];
  L.~Randall and R.~Sundrum,
  Phys.\ Rev.\ Lett.\  {\bf 83}, 3370 (1999)
  [hep-ph/9905221];
  T.~Appelquist, H.~C.~Cheng and B.~A.~Dobrescu,
  Phys.\ Rev.\ D {\bf 64}, 035002 (2001)
  [hep-ph/0012100].

\bibitem{flavor}
For some early work on this subject in the RS model, see, for example, 
  K.~Agashe, G.~Perez and A.~Soni,
  Phys.\ Rev.\ D {\bf 71}, 016002 (2005)
  [hep-ph/0408134];
  S.~J.~Huber,
  Nucl.\ Phys.\ B {\bf 666}, 269 (2003)
  [hep-ph/0303183];
  A.~L.~Fitzpatrick, G.~Perez and L.~Randall,
  Phys.\ Rev.\ Lett.\  {\bf 100}, 171604 (2008)
  [arXiv:0710.1869 [hep-ph]].
 
\bibitem{ued2}
 G.~Servant and T.~M.~P.~Tait,
  Nucl.\ Phys.\ B {\bf 650}, 391 (2003)
  [hep-ph/0206071];
  H.~C.~Cheng, J.~L.~Feng and K.~T.~Matchev,
  Phys.\ Rev.\ Lett.\  {\bf 89}, 211301 (2002)
  [hep-ph/0207125].

\bibitem{morrissey}
For earlier related work employing a modified warped RS setup, see 
  K.~L.~McDonald and D.~E.~Morrissey,
  JHEP {\bf 1005}, 056 (2010)
  [arXiv:1002.3361 [hep-ph]] and 
  JHEP {\bf 1102}, 087 (2011)
  [arXiv:1010.5999 [hep-ph]].
  
  \bibitem{keith}
See, for example,   
  K.~R.~Dienes and B.~Thomas,
  Phys.\ Rev.\ D {\bf 85}, 083523 (2012)
  [arXiv:1106.4546 [hep-ph]] 
 and  
  Phys.\ Rev.\ D {\bf 85}, 083524 (2012)
  [arXiv:1107.0721 [hep-ph]] and subsequent works.
 

 \bibitem{vectorportal} 
 There has been a huge amount of work on this subject; see, for example, 
  D.~Feldman, B.~Kors and P.~Nath,
  Phys.\ Rev.\ D {\bf 75}, 023503 (2007)
  [hep-ph/0610133];
  D.~Feldman, Z.~Liu and P.~Nath,
  Phys.\ Rev.\ D {\bf 75}, 115001 (2007)
  [hep-ph/0702123 [HEP-PH]].;
  M.~Pospelov, A.~Ritz and M.~B.~Voloshin,
  Phys.\ Lett.\ B {\bf 662}, 53 (2008)
  [arXiv:0711.4866 [hep-ph]];
  M.~Pospelov,
  Phys.\ Rev.\ D {\bf 80}, 095002 (2009)
  [arXiv:0811.1030 [hep-ph]]; 
  H.~Davoudiasl, H.~S.~Lee and W.~J.~Marciano,
  Phys.\ Rev.\ Lett.\  {\bf 109}, 031802 (2012)
  [arXiv:1205.2709 [hep-ph]] and 
  Phys.\ Rev.\ D {\bf 85}, 115019 (2012)
  doi:10.1103/PhysRevD.85.115019
  [arXiv:1203.2947 [hep-ph]];
  R.~Essig {\it et al.},
  arXiv:1311.0029 [hep-ph];
  E.~Izaguirre, G.~Krnjaic, P.~Schuster and N.~Toro,
  Phys.\ Rev.\ Lett.\  {\bf 115}, no. 25, 251301 (2015)
  [arXiv:1505.00011 [hep-ph]];
 For a general overview and introduction to this framework, see  
  D.~Curtin, R.~Essig, S.~Gori and J.~Shelton,
  JHEP {\bf 1502}, 157 (2015)
  [arXiv:1412.0018 [hep-ph]].
 
 \bibitem{KM}
  B.~Holdom,
  Phys.\ Lett.\  {\bf 166B}, 196 (1986) and
  Phys.\ Lett.\ B {\bf 178}, 65 (1986); 
  K.~R.~Dienes, C.~F.~Kolda and J.~March-Russell,
  Nucl.\ Phys.\ B {\bf 492}, 104 (1997)
  [hep-ph/9610479];
  F.~Del Aguila,
  Acta Phys.\ Polon.\ B {\bf 25}, 1317 (1994)
  [hep-ph/9404323];
  K.~S.~Babu, C.~F.~Kolda and J.~March-Russell,
  Phys.\ Rev.\ D {\bf 54}, 4635 (1996)
  [hep-ph/9603212];
  T.~G.~Rizzo,
  Phys.\ Rev.\ D {\bf 59}, 015020 (1998)
  [hep-ph/9806397].
  
 \bibitem{higgsportal}
 See, for example, 
  B.~Patt and F.~Wilczek,
  hep-ph/0605188;
  S.~Gopalakrishna, S.~Jung and J.~D.~Wells,
  Phys.\ Rev.\ D {\bf 78}, 055002 (2008)
  [arXiv:0801.3456 [hep-ph]];
  J.~D.~Clarke, R.~Foot and R.~R.~Volkas,
  JHEP {\bf 1402}, 123 (2014)
  [arXiv:1310.8042 [hep-ph]];
  J.~Liu, X.~P.~Wang and F.~Yu,
  JHEP {\bf 1706}, 077 (2017)
  [arXiv:1704.00730 [hep-ph]];
L.~Darmé, S.~Rao and L.~Roszkowski,
  arXiv:1710.08430 [hep-ph];
  J.~L.~Feng, I.~Galon, F.~Kling and S.~Trojanowski,
  arXiv:1710.09387 [hep-ph];
 
 \bibitem{blkts}
  G.~R.~Dvali, G.~Gabadadze and M.~A.~Shifman,
  Phys.\ Lett.\ B {\bf 497}, 271 (2001)
  [hep-th/0010071];
  M.~Carena, E.~Ponton, T.~M.~P.~Tait and C.~E.~M.~Wagner,
  Phys.\ Rev.\ D {\bf 67}, 096006 (2003)
  [hep-ph/0212307];
  M.~Carena, T.~M.~P.~Tait and C.~E.~M.~Wagner,
  Acta Phys.\ Polon.\ B {\bf 33}, 2355 (2002)
  [hep-ph/0207056];
  F.~del Aguila, M.~Perez-Victoria and J.~Santiago,
  Acta Phys.\ Polon.\ B {\bf 34}, 5511 (2003)
  [hep-ph/0310353] and 
  JHEP {\bf 0302}, 051 (2003)
  [hep-th/0302023];
  H.~Davoudiasl, J.~L.~Hewett and T.~G.~Rizzo,
  Phys.\ Rev.\ D {\bf 68}, 045002 (2003)
  [hep-ph/0212279] and 
  JHEP {\bf 0308}, 034 (2003)
  [hep-ph/0305086].

\bibitem{stuff}
See, for example, the last paper in Ref.\cite{vectorportal}. 
 
\bibitem{Heeck:2011md} 
  J.~Heeck and W.~Rodejohann,
  Phys.\ Lett.\ B {\bf 705}, 369 (2011)
  [arXiv:1109.1508 [hep-ph]].

\bibitem{pdg}
C. Patrignani et al. (Particle Data Group), Chin. Phys. C, 40, 100001 (2016) and 2017 update.
 
\bibitem{Ade:2015xua} 
  P.~A.~R.~Ade {\it et al.} [Planck Collaboration],
  Astron.\ Astrophys.\  {\bf 594}, A13 (2016)
  [arXiv:1502.01589 [astro-ph.CO]].

\bibitem{Liu:2016cnk} 
  H.~Liu, T.~R.~Slatyer and J.~Zavala,
  Phys.\ Rev.\ D {\bf 94}, no. 6, 063507 (2016)
  [arXiv:1604.02457 [astro-ph.CO]].
  See also, 
  M.~Dutra, M.~Lindner, S.~Profumo, F.~S.~Queiroz, W.~Rodejohann and C.~Siqueira,
  arXiv:1801.05447 [hep-ph].
  
\bibitem{Steigman:2015hda} 
  G.~Steigman,
  Phys.\ Rev.\ D {\bf 91}, no. 8, 083538 (2015)
  [arXiv:1502.01884 [astro-ph.CO]].
  
  
\bibitem{pair}
See, for example, 
  J.~Berger, K.~Jedamzik and D.~G.~E.~Walker,
  JCAP {\bf 1611}, 032 (2016)
  [arXiv:1605.07195 [hep-ph]];
  A.~Fradette, M.~Pospelov, J.~Pradler and A.~Ritz,
  Phys.\ Rev.\ D {\bf 90}, no. 3, 035022 (2014)
  [arXiv:1407.0993 [hep-ph]].

\bibitem{wip}
  T.~G.~Rizzo,
  arXiv:1805.08150 [hep-ph].
 

\bibitem{Hagiwara:2017zod} 
  K.~Hagiwara, A.~Keshavarzi, A.~D.~Martin, D.~Nomura and T.~Teubner,
  Nucl.\ Part.\ Phys.\ Proc.\  {\bf 287-288}, 33 (2017).

\bibitem{peskin}
M. Peskin and T. Takeuchi,
  Phys.\ Rev.\ D.\ {\bf 46}, 381 (1992)

\bibitem{maksymyk}
I. Maksymyk, C.P. Burgess, D. London,
  Phys.\ Rev.\ D.\ {\bf 50}, 529 (1994)

\bibitem{LDMX}
For more information about the LDMX experiment, see ~\url{https://confluence.slac.stanford.edu/display/MME/Light+Dark+Matter+Experiment} 
and also 
  T.~Raubenheimer {\it et al.},
  arXiv:1801.07867 [physics.acc-ph].


\bibitem{Mans:2017vej} 
  J.~Mans [LDMX Collaboration],
  EPJ Web Conf.\  {\bf 142}, 01020 (2017).

\bibitem{Bjorken:2009mm} 
  J.~D.~Bjorken, R.~Essig, P.~Schuster and N.~Toro,
  Phys.\ Rev.\ D {\bf 80}, 075018 (2009)
  [arXiv:0906.0580 [hep-ph]].
 
 \bibitem{Belle}
  G.~Inguglia,
  PoS DIS {\bf 2016}, 263 (2016)
  [arXiv:1607.02089 [hep-ex]].
 
\bibitem{Liu:2017jzn} 
  Z.~C.~Liu, C.~X.~Yue and Y.~C.~Guo,
  arXiv:1703.00153 [hep-ph].
 
\bibitem{mono}
  M.~Aaboud {\it et al.} [ATLAS Collaboration],
  arXiv:1711.03301 [hep-ex];
  A.~M.~Sirunyan {\it et al.} [CMS Collaboration],
  arXiv:1712.02345 [hep-ex].

\bibitem{Chen:2016tdz} 
  C.~S.~Chen, G.~L.~Lin, Y.~H.~Lin and F.~Xu,
  Int.\ J.\ Mod.\ Phys.\ A {\bf 32}, no. 31, 1750178 (2017)
  [arXiv:1609.07198 [hep-ph]]. 
See also, 
  T.~Emken, C.~Kouvaris and I.~M.~Shoemaker,
  Phys.\ Rev.\ D {\bf 96}, no. 1, 015018 (2017)
  [arXiv:1702.07750 [hep-ph]]  and 
  M.~J.~Dolan, F.~Kahlhoefer and C.~McCabe,
  arXiv:1711.09906 [hep-ph].

\bibitem{Essig:2017kqs} 
  R.~Essig, T.~Volansky and T.~T.~Yu,
  Phys.\ Rev.\ D {\bf 96}, no. 4, 043017 (2017)
  [arXiv:1703.00910 [hep-ph]].
 
\bibitem{Essig:2015cda} 
  R.~Essig, M.~Fernandez-Serra, J.~Mardon, A.~Soto, T.~Volansky and T.~T.~Yu,
  JHEP {\bf 1605}, 046 (2016)
  [arXiv:1509.01598 [hep-ph]].

  
\bibitem{Berlin:2014tja} 
See, for example, 
  A.~Berlin, D.~Hooper and S.~D.~McDermott,
  Phys.\ Rev.\ D {\bf 89}, no. 11, 115022 (2014)
  [arXiv:1404.0022 [hep-ph]].


\bibitem{5d}
  A.~Muck, A.~Pilaftsis and R.~Ruckl,
  Phys.\ Rev.\ D {\bf 65}, 085037 (2002)
  [hep-ph/0110391];
  T.~Flacke, A.~Menon and D.~J.~Phalen,
  Phys.\ Rev.\ D {\bf 79}, 056009 (2009)
  [arXiv:0811.1598 [hep-ph]].

\bibitem{Davoli:2017swj} 
See, for example,
  A.~Davoli, A.~De Simone, T.~Jacques and V.~Sanz,
  JHEP {\bf 1711}, 025 (2017)
  [arXiv:1706.08985 [hep-ph]].

\bibitem{nikita}
See, for example, 
  N.~Blinov, E.~Izaguirre and B.~Shuve,
  Phys.\ Rev.\ D {\bf 97}, no. 1, 015009 (2018)
  doi:10.1103/PhysRevD.97.015009
  [arXiv:1710.07635 [hep-ph]].


\bibitem{Feng:2017drg} 
We follow the analysis for light dark matter as presented in  J.~L.~Feng and J.~Smolinsky,
  Phys.\ Rev.\ D {\bf 96}, no. 9, 095022 (2017)
  [arXiv:1707.03835 [hep-ph]] .
See also, 
  B.~Li and Y.~F.~Zhou,
  Commun.\ Theor.\ Phys.\  {\bf 64}, no. 1, 119 (2015)
  [arXiv:1503.08281 [hep-ph]] 
and  
  B.~Li and Y.~F.~Zhou,
  Commun.\ Theor.\ Phys.\  {\bf 64}, no. 1, 119 (2015)
  [arXiv:1503.08281 [hep-ph]].

\bibitem{Slatyer:2016qyl} 
  T.~R.~Slatyer and C.~L.~Wu,
  Phys.\ Rev.\ D {\bf 95}, no. 2, 023010 (2017)
  [arXiv:1610.06933 [astro-ph.CO]].


\bibitem{21club}
See, for example, 
  H.~Liu and T.~R.~Slatyer,
  arXiv:1803.09739 [astro-ph.CO];
  S.~Clark, B.~Dutta, Y.~Gao, Y.~Z.~Ma and L.~E.~Strigari,
  arXiv:1803.09390 [astro-ph.HE];
  A.~Mitridate and A.~Podo,
  arXiv:1803.11169 [hep-ph].




\end{thebibliography}
\end{document}